\documentclass[aps,prc,10pt,amsmath,amssymb,twocolumn,floatfix,nofootinbib,
showpacs,noeprint,superscriptaddress]{revtex4-1}

\usepackage{amssymb,amsthm,amsmath,amstext,amsbsy}
\usepackage{bbm}
\usepackage{url}
\usepackage{nicefrac}
\usepackage{graphicx}
\usepackage{hyperref}
\usepackage{leftidx}
\usepackage{xspace}
\usepackage{array}
\usepackage{textgreek}
\DeclareMathSymbol{\NS}{\mathord}{AMSb}{"4E}

\newcommand{\ket}[1]{\ensuremath{\,|{#1}\rangle}}

\newcommand{\matrixe}[3]{\ensuremath{\langle{#1}|\,{#2}\,|{#3}\rangle}}

\newcommand{\op}[1]{\ensuremath{#1}}
\newcommand{\adj}[1]{\ensuremath{{{#1}}^{\dag}}}

\newcommand{\aO}{\ensuremath{\op{a}}}

\newcommand{\etaO}{\ensuremath{\op{\eta}}}
\newcommand{\aaO}{\ensuremath{\adj{\op{a}}}}

\newcommand{\HO}{\ensuremath{\op{H}}}

\newcommand{\totd}[2]{\ensuremath{ \frac{d {#1}} {d {#2}} }}

\begin{document}
\title{The Magnus expansion and the in-medium similarity renormalization group}
\author{T.~D.~Morris}
\email[E-mail:~]{morrist@nscl.msu.edu}
\author{N.~Parzuchowski}
\email[E-mail:~]{parzuchowski@frib.msu.edu}
\author{S.~K.~Bogner}
\email[E-mail:~]{bogner@nscl.msu.edu}
\affiliation{National Superconducting Cyclotron Laboratory
and Department of Physics and Astronomy, Michigan State University,
East Lansing, Michigan 48824, USA}
\date{\today}
\begin{abstract}
We present an improved variant of the in-medium similarity renormalization group (IM-SRG) based on the Magnus expansion. In the new formulation, one solves flow equations for the anti-hermitian operator that, upon exponentiation, yields the unitary transformation of the IM-SRG. The resulting flow equations can be solved using a first-order Euler method without any loss of accuracy, resulting in substantial memory savings and modest computational speedups. Since one obtains the unitary transformation directly, the transformation of additional operators beyond the Hamiltonian can be accomplished with little additional cost, in sharp contrast to the standard formulation of the IM-SRG. Ground state calculations of the homogeneous electron gas (HEG) and $^{16}$O nucleus are used as test beds to illustrate the efficacy of the Magnus expansion.   
\end{abstract}
\maketitle
\clearpage
\section{Introduction\label{sec:intro}}
The quest to predict and understand the properties of exotic nuclei starting from the underlying nuclear forces represents a cornerstone of modern nuclear theory. Already for stable nuclei, there are computational and theoretical challenges that make the ab-initio description of nuclear structure quite difficult. Nevertheless, tremendous progress has been made over the past two decades, where it is now possible to perform quasi-exact calculations including three-nucleon interactions of nuclei up through Carbon or so in Quantum Monte Carlo 
(QMC) and No-Core Shell Model (NCSM) 
calculations, and $N=Z$ nuclei up through $^{28}$Si in lattice effective field theory with Euclidean time projection~\cite{Lovato2014,Barrett2013,Lahde:2013uqa}.

Since exact methods scale unfavorably with system size, it is necessary to develop approximate, but systematically improvable methods to extend the reach of ab-initio theory beyond light nuclei. Over the past decade, Coupled Cluster (CC) theory, Self-Consistent Green's Functions (SCGF), Auxiliary Field Diffusion Monte Carlo (AFDMC), and the In-Medium Similarity Renormalization Group (IM-SRG) have been successfully applied to calculate properties of selected medium mass nuclei and infinite nuclear matter~\cite{hagen2013c,Soma:2013,carbone2013, Hergert:2014iaa,Tsukiyama:2010rj,Roth:2012qf,Gandolfi:2014ewa}.  Early applications of these methods were limited primarily to ground state properties of stable nuclei near shell closures with two-nucleon forces only.  In recent years, however, substantial progress has been made on including three-nucleon forces~\cite{Hagen:2007zc,Soma:2013,Roth:2012qf, Hergert:2012nb}, targeting excited states and observables besides energy~\cite{Ekstrom:2014iya,Jansen:2012ey}, and moving into the more challenging terrain of open-shell and unstable nuclei~\cite{Tsukiyama:2012sm,Jansen:2014qxa,Bogner:2014baa,Soma:2012zd,Hergert:2013uja}. 

The IM-SRG is a particularly appealing method due to its flexibility to target ground and excited state properties for closed- and open-shell systems. As discussed in Sec.~\ref{sec:formalism}, the essence of the IM-SRG is to perform a continuous unitary transformation on the Hamiltonian (and all other observables of interest) to drive it to a diagonal or block-diagonal form.  The transformation is implemented by solving a coupled set of flow equations for the matrix elements of the Hamiltonian and any other operators of interest
\begin{eqnarray}
\label{eq:srg1}
H(s) = U^{\dagger}(s)HU(s)\quad&\Leftrightarrow&\quad \frac{dH(s)}{ds}= [\eta(s),H(s)] \nonumber\\
O(s) = U^{\dagger}(s)OU(s)\quad&\Leftrightarrow&\quad \frac{dO(s)}{ds}= [\eta(s),O(s)]\,,\nonumber\\
\end{eqnarray}
where $s$ is a continuous flow parameter, and the choice of the generator $\eta(s)\equiv\frac{dU^{\dagger}}{ds}U$ implicitly defines the transformation $U(s)$.  Despite the flexibility to tailor $\eta$ to a wide range of problems and the modest computational scaling with system size, the formulation in Eq.~\ref{eq:srg1} suffers from the following difficulties: 
\begin{itemize}
\item The coupled ODEs can become stiff for certain choices of generator and/or for systems with strong correlations.
\item The numerical integration of Eq.~\ref{eq:srg1} requires a high-order ODE solver to accurately preserve the eigenvalues of the evolved Hamiltonian. The use of a high-order solver consumes a large amount of memory since multiple copies of the solution vector (e.g., 15-20 for the predictor-corrector solver of Shampine and Gordon~\cite{ShampineGordon}) need to be stored at each time step.  
\item For each additional observable of interest, the number of coupled ODEs that need to be solved is roughly doubled, assuming a comparable level of truncation for the evolved operator as the Hamiltonian. Moreover, the flow equations for the additional observable(s) can exacerbate the problems with stiff ODEs, since the time scales for the operator evolution may be very different from those of the Hamiltonian. 
\end{itemize}

In the present paper, we will demonstrate how these difficulties can be circumvented\footnote{See also the recent Driven Similarity Renormalization Group (DSRG) approach of Ref.~\cite{Evangelista}, where the problem is recast so that one solves non-linear amplitude equations instead of flow equations.}by using the Magnus expansion to recast Eq.~\ref{eq:srg1} as a flow equation for the operator $\Omega(s)$, where $U(s) = e^{\Omega(s)}$. The unitary operator $U(s)$ is subsequently used to transform the Hamiltonian and any other operators of interest via the Baker-Cambell-Hausdorff (BCH) formula.  We will show that in the Magnus expansion formulation, one can use a naive first-order forward Euler method to solve the flow equations for $\Omega(s)$ without any loss of accuracy. This provides a substantial reduction in memory consumption, and allows operators beside the Hamiltonian to be evolved with little additional cost, in sharp contrast to the original formulation of the IM-SRG based on direct integration of Eq.~\ref{eq:srg1}. 

The remainder of the paper is organized as follows: In Section~\ref{sec:formalism}, we review the basic formalism of the SRG and illustrate how the Magnus expansion can be used to make its numerical implementation more efficient for a schematic model.  In Section~\ref{sec:implementation}, we give some implementation details of our IM-SRG and Magnus expansion calculations of the homogeneous electron gas (HEG) and $^{16}$O nucleus.  Results are presented in Section~\ref{sec:results}, and conclusions are presented in Section~\ref{sec:conclusions}.

\section{\label{sec:formalism}Formalism}
\subsection{\label{sec:SRG}SRG}
The similarity renormalization group consists of a continuous sequence of unitary transformations that gradually suppress
off-diagonal matrix elements, driving the Hamiltonian towards a
band- or block-diagonal form~\cite{Glazek:1993rc,wegner94,Bogner:2006pc,Binder:2012uq}.  Writing the transformed Hamiltonian as

\begin{align}
 H(s)&=U(s)HU^\dagger(s)\equiv H^d(s)+H^{od}(s),\label{eq:ham_uni_trans}
\end{align}
where $H^{d}(s)$ and $H^{od}(s)$ are the arbitrarily defined ``diagonal'' and ``off-diagonal'' parts of the Hamiltonian, the evolution with the continuous flow parameter $s$
is given by 
\begin{align}
\label{eq:srg}
 \frac{dH(s)}{ds}&=[\eta(s),H(s)],
\end{align}
where the $\eta(s)\equiv U(s)dU^{\dagger}(s)/ds$ is the (anti-hermitian) generator of the transformation. The choice of the generator first suggested by Wegner,
\begin{align}
 \eta(s)&=[H^d(s),H(s)]=[H^d(s),H^{od}(s)],\label{eq:eta_weg4}
\end{align}
guarantees
\begin{align}
 \frac{d}{ds}Tr\left((H^{od})^2\right)&=2Tr(\eta^2)=-2Tr(\eta^\dagger \eta)\le 0,
\end{align}
which demonstrates 
that the strength of $H^{od}$ decays with increasing
$s$\cite{wegner94}. By analyzing the flow equations in the eigenbasis of $H^{d}(s)$ and defining $H^d_{ii}(s)\equiv \epsilon_i$, one can show that the Wegner generator gives a super-exponential decay of the off-diagonal matrix elements

\begin{equation}
\label{eq:wegnerdecay}
H^{od}_{ij}(s)\sim e^{-s(\epsilon_i-\epsilon_j)^2}H^{od}_{ij}(0)\,.
\end{equation}

The SRG evolution with the Wegner generator closely resembles the conventional Wilsonian RG, since matrix elements between widely-separated energy scales are eliminated before moving inwards towards the diagonal.  The Wegner generator is numerically very stable, but the different rates of decay for off-diagonal matrix elements can lead to stiff ODEs.  To avoid this, two alternative classes of generators are commonly used in nuclear applications. The first alternative was proposed by White in Ref.~\cite{White:2002fk}
\begin{equation}
\label{eq:White}
\eta_{ij}(s) = \frac{H^{od}_{ij}(s)}{\epsilon_i - \epsilon_j}\,,
\end{equation}
which leads to a uniform suppression of off-diagonal matrix elements
\begin{equation}
H^{od}_{ij}(s)\sim e^{-s}H^{od}_{ij}(0)\,.
\end{equation}
The White generator is numerically very efficient for well-behaved problems, though it can become unstable when the energy denominator in Eq.~\ref{eq:White} becomes too small. In Ref.~\cite{Hergert:2014iaa}, the so-called ``imaginary time'' generator was proposed as a compromise between the White and Wegner generators
\begin{equation}
\eta_{ij}(s) = sign\bigl(\epsilon_i-\epsilon_j\bigr)H^{od}_{ij}(s)\,,
\end{equation} 
where the leading behavior of the off-diagonal matrix elements is
\begin{equation}
H^{od}_{ij}(s) \sim e^{-s|\epsilon_i-\epsilon_j|}H^{od}_{ij}(0)\,.
\end{equation}
In the present paper, all of our IM-SRG calculations were done using the White generator. However, we stress that the computational benefits of the Magnus expansion carry over irrespective of the specific choice of generator.  

\subsection{\label{floweqns}In-medium Evolution}
Until recently, most SRG applications to nuclear interactions have been carried out
in free space to ``soften'' two- and three-nucleon interactions to be used as input for ab-initio
calculations~\cite{Bogner:2007rx,Bogner:2009bt,Binder:2012uq}. The free-space evolution is convenient, as it does not have to be performed for each different nucleus or nuclear matter density. However, it is
necessary to consistently evolve three-nucleon (and possibly higher) interactions to be
able to soften the interactions significantly and maintain approximate
$s$-independence of $A \geqslant 3$ observables. The consistent SRG evolution of three-nucleon operators represents a significant technical challenge that has only recently been solved in recent years~\cite{Jurgenson:2009qs,Hebeler:2012pr,Wendt:2013bla}. 

An interesting alternative is to perform the SRG evolution in-medium (IM-SRG) for each $A$-body system of interest~\cite{Bogner:2009bt,Tsukiyama:2010rj}.  Unlike the free-space evolution, the IM-SRG has the appealing feature that one can approximately evolve
$3,...,A$-body operators using only two-body machinery by 
normal-ordering with respect to an $A$-body reference state. Moreover, with a suitable definition of the off-diagonal part of the Hamiltonian to be driven to zero, the IM-SRG can be used as an ab-initio method in and of itself, rather than simply to soften the Hamiltonian as in the free-space SRG.

Starting from a general second-quantized Hamiltonian with two- and three-body interactions,
\begin{align}
 \hat{H}&=\sum_{qr}T_{qr}a_q^\dagger a_r +\frac{1}{2!^2}\sum_{qrst}V^{(2)}_{qrst}a_q^\dagger a_r^\dagger a_ta_s \nonumber \\
&+\frac{1}{3!^2}\sum_{qrstuv}V^{(3)}_{qrstuv}a_q^\dagger a_r^\dagger a_s^\dagger a_va_ua_t +\cdots \label{eq:initial_hamiltonian}
\end{align}
all operators can be normal-ordered with respect to a finite-density 
Fermi vacuum $|\Phi\rangle$ (e.g., the Hartree-Fock ground
state), as opposed to the zero-particle vacuum\footnote{In the present work, we restrict our attention to single reference (i.e., closed-shell) systems for which a single Slater determinant provides a reasonable starting point. See Refs.~\cite{Tsukiyama:2012sm,Bogner:2014baa,Hergert:2013uja} for extensions of the IM-SRG to open-shell systems.}. Wick's theorem can then be
used to {\it exactly} write $H$ as
\begin{align}\label{eq:Hno}
  \HO &= E + \sum_{qr}f_{qr}:\aaO_{q}\aO_{r}: + \frac{1}{4}\sum_{qrst}\Gamma_{qrst}:\aaO_{q}\aaO_{r}\aO_{t}\aO_{s}:\notag\\
   	  &\hphantom{=}+ \frac{1}{36}\sum_{qrstuv}W_{qrstuv}:\aaO_{q}\aaO_{r}\aaO_{s}\aO_{v}\aO_{u}\aO_{t}:\,,
\end{align}
where strings of normal-ordered operators obey the following relation.
\begin{equation}
  \matrixe{\Phi}{:\aaO_{q}\ldots\aO_{r}:}{\Phi} = 0\,,
\end{equation}
and the terms in Eq.~\eqref{eq:Hno} are given by
\begin{align}
\label{eq:nordering}
  E &= \sum_{q}T_{qq}n_{q}
  		\hphantom{=}+ \frac{1}{2}\sum_{qr}\!V^{(2)}_{qrqr}n_{q}n_{r}\notag\\
  		&\hphantom{=}+ \frac{1}{6}\sum_{qrs}V^{(3)}_{qrsqrs}n_{q}n_{r}n_{s}\,,\\
  f_{qr} &= T_{qr}
  		+ \sum_{s}\!V^{(2)}_{qsrs}n_{s}
  		+\frac{1}{2}\sum_{st}V^{(3)}_{qstrst}n_{s}n_{t}\,,\label{eq:f}		\\
  \Gamma_{qrst} &= \!V^{(2)}_{qrst} + \sum_{u}V^{(3)}_{qrustu}n_{u}\,,\\
  W_{qrstuv}&=V^{(3)}_{qrstuv}\,.
\end{align}
Here, the initial $n$-body interactions are denoted by $V^{(n)}$, and
$n_q=\theta(\epsilon_{\rm F}-\epsilon_q)$ are occupation numbers in
the reference state $|\Phi \rangle$, with Fermi energy $\epsilon_{\rm
F}$. It is evident that
the normal-ordered $0$-, $1$-, and $2$-body terms include contributions
from the three-body interaction $V^{(3)}$ through sums over the
occupied single-particle states in the reference state
$|\Phi\rangle$. Neglecting the residual three-body interaction leads to the normal-ordered two-body approximation (NO2B), which has been shown to be an excellent approximation in many nuclear systems~\cite{Roth:2012qf,Binder:2012uq,Hagen:2007zc}. Truncating the in-medium SRG equations to
normal-ordered two-body operators, which we denote by IM-SRG(2),
will approximately evolve induced three- and higher-body interactions
through the nucleus-dependent 0-, 1-, and 2-body terms.

Using Wick's theorem to evaluate Eq.~\ref{eq:srg} with
$H(s) = E_0(s) + f(s)+ \Gamma(s)$ and $\eta(s) = \eta^{(1)}(s) + \eta^{(2)}(s)$
truncated to normal-ordered two-body operators, one obtains the
coupled IM-SRG(2) flow equations

\begin{align}
  \label{eq:0bflow}
  \totd{E}{s}&=\sum_{qr}\eta_{qr}f_{rq}(n_{q}-n_{r})
                                        +\frac{1}{2}\sum_{qrst}\eta_{qrst}\Gamma_{stqr}n_{q}n_{r}\bar{n}_{s}\bar{n}_{t}\,,\\
  \label{eq:1bflow}
  \totd{f_{qr}}{s}&=  \sum_{s}(1+P_{qr})\eta_{qs}f_{sr} \notag\\ &\quad+\sum_{st}(n_s-n_t)(\eta_{st}\Gamma_{tqsr}-f_{st}\eta_{tqsr}) \notag\\ 
  &\quad +\sum_{stu}(n_sn_t\bar{n}_u+\bar{n}_s\bar{n}_tn_u) (1+P_{qr})\eta_{uqst}\Gamma_{stur}\\
  \label{eq:2bflow}
 \totd{\Gamma_{qrst}}{s}
                                &=  \sum_{u}\left\{ 
    (1-P_{qr})(\eta_{qu}\Gamma_{urst}-f_{qu}\eta_{urst} )\right\} \notag\\ 
    &\quad-\sum_{u}\left\{(1-P_{st})(\eta_{us}\Gamma_{qrut}-f_{us}\eta_{qrut} )
    \right\}\notag \\
  &\quad+ \frac{1}{2}\sum_{uv}(1-n_u-n_v)(\eta_{qruv}\Gamma_{uvst}-\Gamma_{qruv}\eta_{uvst})
    \notag\\
  &\quad-\sum_{uv}(n_u-n_v) (1-P_{qr})(1-P_{st})\eta_{vrut}\Gamma_{uqvs}\,,
                                                                                                                                                    \end{align}
where $\bar{n}_r\equiv 1-n_r$ and the $s$-dependence has been suppressed for brevity. 

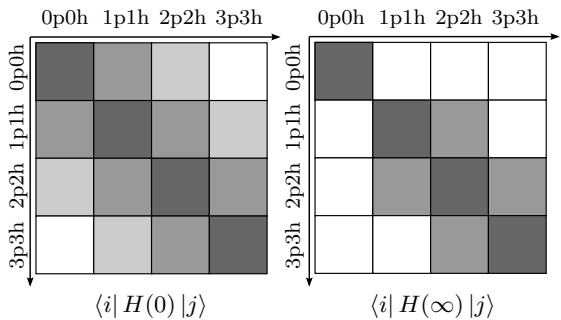
\begin{figure}[t]
\setlength{\unitlength}{0.85\columnwidth}
  \begin{center}
  \begin{picture}(1.0000,0.5500)
   \put(0.0350,0.0450){\includegraphics[width=0.46\unitlength]{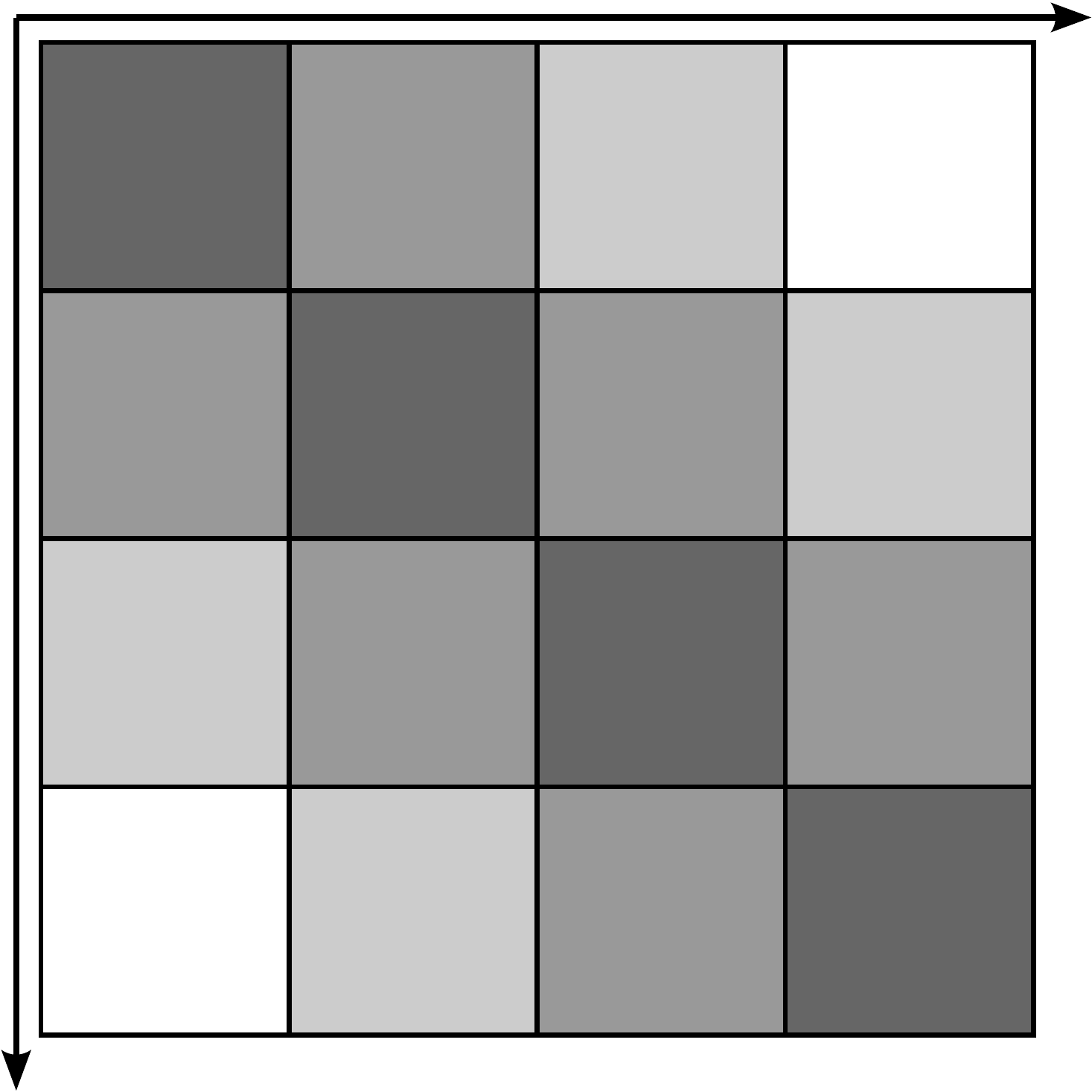}}
   \put(0.5400,0.0450){\includegraphics[width=0.46\unitlength]{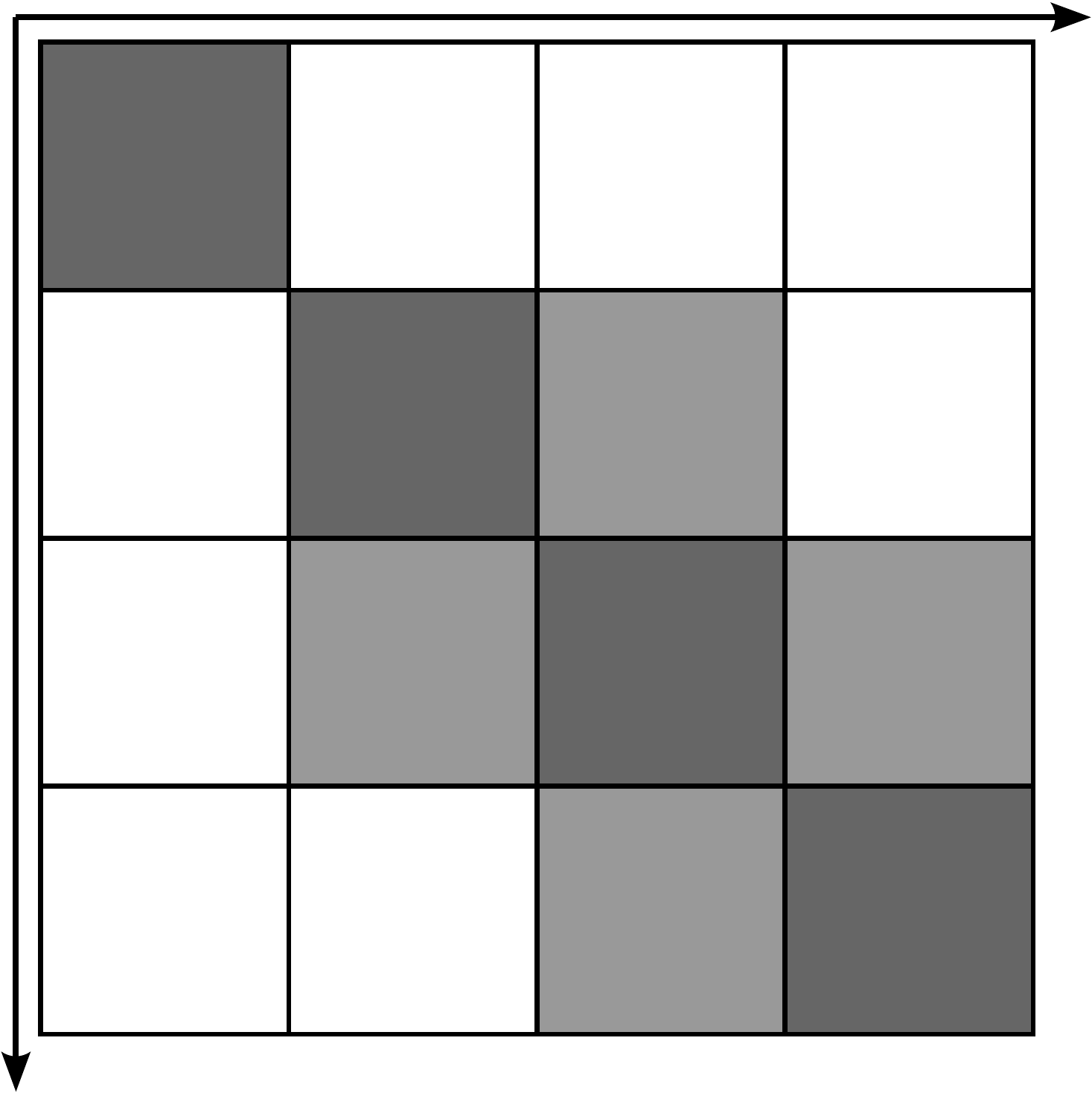}}
   \put(0.0100,0.0000){\parbox{0.5\unitlength}{\centering$\matrixe{i}{\HO(0)}{j}$}}
   \put(0.5200,0.0000){\parbox{0.5\unitlength}{\centering$\matrixe{i}{\HO(\infty)}{j}$}}
   
   \put(0.0500,0.5200){\parbox{0.11\unitlength}{\centering\footnotesize0p0h}}
   \put(0.1600,0.5200){\parbox{0.11\unitlength}{\centering\footnotesize1p1h}}
   \put(0.2630,0.5200){\parbox{0.11\unitlength}{\centering\footnotesize2p2h}}
   \put(0.3650,0.5200){\parbox{0.11\unitlength}{\centering\footnotesize3p3h}}
   \put(0.5500,0.5200){\parbox{0.11\unitlength}{\centering\footnotesize0p0h}}
   \put(0.6600,0.5200){\parbox{0.11\unitlength}{\centering\footnotesize1p1h}}
   \put(0.7630,0.5200){\parbox{0.11\unitlength}{\centering\footnotesize2p2h}}
   \put(0.8650,0.5200){\parbox{0.11\unitlength}{\centering\footnotesize3p3h}}
   \put(-0.0350000,0.4320){\parbox{0.11\unitlength}{\rotatebox{90}{\centering\footnotesize0p0h}}}
   \put(-0.035000,0.3235){\parbox{0.11\unitlength}{\rotatebox{90}{\centering\footnotesize1p1h}}}
   \put(-0.035000,0.2175){\parbox{0.11\unitlength}{\rotatebox{90}{\centering\footnotesize2p2h}}}
   \put(-0.035000,0.1100){\parbox{0.11\unitlength}{\rotatebox{90}{\centering\footnotesize3p3h}}}

   \put(0.465000,0.4320){\parbox{0.11\unitlength}{\rotatebox{90}{\centering\footnotesize0p0h}}}
   \put(0.465000,0.3235){\parbox{0.11\unitlength}{\rotatebox{90}{\centering\footnotesize1p1h}}}
   \put(0.465000,0.2175){\parbox{0.11\unitlength}{\rotatebox{90}{\centering\footnotesize2p2h}}}
   \put(0.465000,0.1100){\parbox{0.11\unitlength}{\rotatebox{90}{\centering\footnotesize3p3h}}}
  \end{picture}
  \end{center}
  \caption{\label{fig:schematic}Schematic representation of the initial and final Hamiltonians, $\HO(0)$ and $\HO(\infty)$, in the many-body Hilbert space spanned by particle-hole excitations of the reference state.}
\end{figure}

For the calculation of the ground state of a closed-shell system in the IM-SRG(2) approximation, it is simple to identify $H^{\rm od} = \{\Gamma_{abij},f_{ai},+h.c\}$, where $a,b$ denote particle (unoccupied) and $i,j$ hole (occupied) single particle states, as the relevant vertices which connect our chosen reference state $\ket{\Phi}$ with higher particle-hole excitations, see Fig.~\ref{fig:schematic}.  By designing a generator to eliminate these terms, one finds that the 0-body term approaches the interacting ground state energy in the limit of large $s$,
\begin{equation}
\lim_{s\rightarrow\infty}E_0(s) = \langle\Phi|H(s)|\Phi\rangle = E_{gs}\,.
\end{equation}
In the present paper we use the White generator~\cite{White:2002fk}, though we deviate slightly from recent implementations that use Epstein-Nesbett energy denominators~\cite{Hergert:2012nb}, opting for the simpler M{\o}ller-Plosset energy denominators 
\begin{align}
  \etaO
  &=\sum_{ai}\frac{f_{ai}}{f_{a}-f_{i}}:\aaO_{a}\aO_{i}:\notag\\&+\frac{1}{4}\sum_{abij}\frac{\Gamma_{abij}}{f_{a}+f_{b}-f_{i}-f_{j}}:\aaO_{a}\aaO_{b}\aO_{j}\aO_{i}:-\text{H.c.}\,,\label{eq:eta}
\end{align}
where $f_{a}=f_{aa}$, etc. The use of M{\o}ller-Plosset denominators has minimal impact on the results of ground state IM-SRG(2) calculations, but it has the virtue of revealing the connection to MBPT. In a subsequent work, this connection will be used to develop approximations that go beyond the IM-SRG(2).

\subsection{\label{sec:magnusform}Magnus Expansion}
IM-SRG calculations typically use ODE solvers based on high-order Runge-Kutta or predictor-corrector methods to solve Eq.~\ref{eq:srg}. The use of a high-order method is essential as the accumulation of time-step errors will destroy the unitary equivalence between $H(s)$ and $H(0)$, even if no truncations are made in the flow equations. State-of-the-art solvers can require the storage of 15-20 copies of the solution vector in memory, which becomes problematic for large model spaces. The problem is exacerbated if one wants to calculate additional observables, roughly doubling the memory requirements  assuming the same NO2B truncation as for the Hamiltonian. Moreover, the additional flow equations for each observable can evolve with rather different timescales than the Hamiltonian, which increases the likelihood of the ODEs becoming stiff. 

To bypass these limitations, we now describe an alternative method to solving Eq. \eqref{eq:srg} using the Magnus expansion~\cite{Magnus54}. In the notation of our present problem, our starting point is the differential equation obeyed by the unitary transformation,
\begin{equation}
\label{eq:matdiff}
\frac{dU(s)}{ds}=-\eta(s) U(s)\,,
\end{equation} 
where $U(0)=1$ and $U^{\dagger}(s)U(s)=U(s)U^{\dagger}(s)=1$. This can be formally integrated and written as the time-ordered exponential
\begin{align}
U(s)&= T_s\bigl\{e^{-\int_0^s \eta(s')ds'}\bigr\}\label{eq:torderexp}\\
&\equiv 1 -\int_{0}^s ds'\eta(s') +\int_0^s ds'\int_0^{s'}ds'' \eta(s')\eta(s'') + \ldots\nonumber\\
\label{eq:torderexp2}
\end{align}
Eq.~\ref{eq:torderexp2} is not very useful in practical calculations since i) there is no guidance on how the series should be truncated, ii) one would need to store $\eta$ for multiple $s$-values, and iii) it is not obvious how to consistently transform the Hamiltonian and other observables in a fully linked, size-extensive manner with the truncated series. 

The Magnus expansion is essentially a statement that, given a few technical requirements on $\eta(s)$,  
a solution of the form 

\begin{equation}
\label{eq:matexp}
U(s)= e^{\Omega(s)}
\end{equation} 
exists, where $\Omega^{\dagger}(s)=-\Omega(s)$ and $\Omega(0)=0$. In most previous applications of the Magnus expansion, one typically expands $\Omega(s)$ in powers of $\eta(s)$ as 
\begin{equation}
\label{eq:magnusexpansion}
\Omega = \sum_{n=1}^\infty \Omega_n\,.
\end{equation}
For issues of convergence and mathematical details, see Refs.~\cite{Blanes2009,feldman84}. Combining this with the formally exact derivative 

\begin{equation}
\label{eq:magnusderiv}
\begin{split}
\frac{d\Omega}{ds}= \sum_{k=0}^\infty \frac{B_k}{k!} ad_{\Omega}^k(\eta)\,\\
ad_{\Omega}^0(\eta)=\eta\\
ad_{\Omega}^k(\eta) = [\Omega,ad_{\Omega}^{k-1}(\eta)]\,,
\end{split}
\end{equation}
where $B_k$ are the Bernoulli numbers and $ad_{\Omega}^k(\eta)$ the recursively defined nested commutators, one can obtain explicit expressions for the $\Omega_n(s)$,  
\begin{eqnarray}
\label{eq:omega_n}
\begin{split}
\Omega_1(s) &= -\int_0^sds_1 \eta(s_1)\\
\Omega_2(s) &= \frac{1}{2}\int_0^s ds_1 \int_0^{s_1} ds_2 [\eta(s_1),\eta(s_2)]\\
\vdots
\end{split}
\end{eqnarray}
As expected, rewriting the time-ordered exponential as a true matrix exponential moves the complications of time ordering into the expression for $\Omega(s)$. 
The utility of the Magnus expansion lies in the fact that, even if $\Omega$ is truncated to low-orders in $\eta$, the resulting transformation in Eq.~\ref{eq:matexp} using the approximate $\Omega$ is unitary, in contrast to any truncated version of Eq.~\ref{eq:torderexp}. 

For large-scale IM-SRG calculations, the expressions in Eq.~\ref{eq:omega_n} are of limited value since they require the storage of $\eta(s)$ over a range of $s$-values.  Therefore, in the present work we instead construct $\Omega(s)$ by numerically integrating Eq.~\ref{eq:magnusderiv}, subject to certain approximations discussed below. The transformed Hamiltonian, and any other operator of interest, can then be constructed by applying the Baker-Cambell-Hausdorff (BCH) formula, 

\begin{eqnarray}
H(s)&= e^{\Omega}H\,e^{-\Omega} = \sum_{k=0}^\infty \frac{1}{k!} ad_{\Omega}^k(H) \label{eq:bch}\\
O(s)&=e^{\Omega}O\,e^{-\Omega} = \sum_{k=0}^\infty \frac{1}{k!} ad_{\Omega}^k(O)\label{eq:bchop}\,.
\label{eq:bchop}
\end{eqnarray}

\begin{figure}[t]

  \includegraphics[width=8cm]{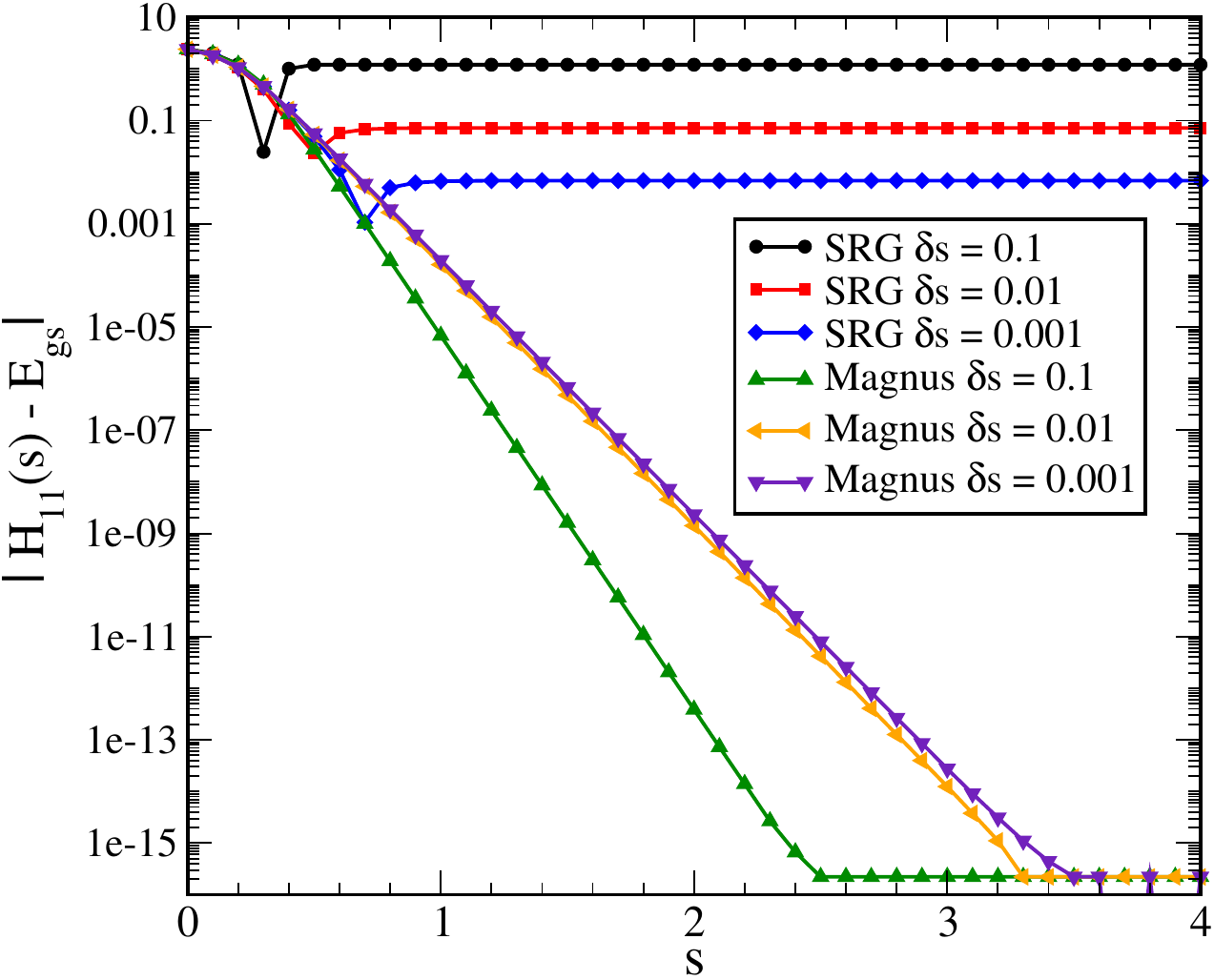}  
  \caption{\label{fig:magconv}(Color online) $|H_{11}(s) - E_{\rm gs}|$ versus $s$ for different Euler step sizes calculated via direct integration of the SRG flow equation, Eq.~\ref{eq:srg}, and using the Magnus expansion, Eqs.~\ref{eq:magnusderiv} and~\ref{eq:bch}.
}
\end{figure}

Before discussing how we truncate Eqs.~\ref{eq:magnusderiv} and~\ref{eq:bch} in practical calculations, it is instructive to study a simple matrix model that can be solved without any truncations. Consider the initial Hamiltonian 
\begin{equation}
\label{eq:2by2}
H = T +V = 
\begin{pmatrix}
1 & 1 \\
1 & -1
\end{pmatrix}\,,
\end{equation}
where the diagonal ``kinetic energy'' term is 
\begin{equation}
T =
\begin{pmatrix}
1 & 0\\
0 & -1
\end{pmatrix}\,.
\end{equation}
Let us now try to diagonalize $H$ using the Wegner generator $\eta(s) =[T,H(s)]$, solving the SRG equations using the Magnus expansion and by direct integration of Eq.~\ref{eq:srg}. Note that for this choice of initial $H$, both $\eta(s)$ and $\Omega(s)$ are real, antisymmetric matrices throughout the flow 
\begin{align}
\eta(s) &=ig_{\eta}(s)\sigma_2\\
\Omega(s) &= ig_{\Omega}(s) \sigma_2\,,
\end{align}
where $\sigma_2$ is the Pauli matrix. Consequently, Eq.~\ref{eq:magnusderiv} terminates at the first term and Eq.~\ref{eq:bch} can be summed up to all orders using the well-known properties of Pauli matrices. Since the large memory footprint of high-order adaptive solvers is the main computational challenge in large-scale SRG calculations, let us instead try to use a naive first-order Euler method to integrate Eqs.~\ref{eq:srg} and~\ref{eq:magnusderiv}. The results are shown in Fig.~\ref{fig:magconv}, where we plot $|H_{11}(s) - E_{gs}|$ -- which should go to zero at large $s$ -- versus $s$ for different Euler step sizes $\delta s$. Unsurprisingly, we see that the direct integration of Eq.~\ref{eq:srg} accumulates large time-step errors, with the plateaus at large $s$ displaying a strong dependence on the Euler step size. The Magnus solution, on the other hand, converges to a final answer at large $s$ that is independent of step size and agrees with the exact result to within machine precision. The insensitivity to the time step size is due to the fact that while each Euler step in Eq.~\ref{eq:magnusderiv} gives an error of order $\mathcal{O}(\delta s^2)$, the exponentiated operator at the end of the evolution is still unitary. This is the primary advantage of the Magnus expansion; by reformulating the problem to solve flow equations for $\Omega(s)$ instead of $H(s)$, one can use a simple first-order Euler method and dramatically reduce memory usage. Once $\Omega(s)$ is in hand, the transformation of $H(s)$ and any other observables of interest immediately follows from Eq.~\ref{eq:bch}. In contrast to the direct integration of Eq.~\ref{eq:srg1}, the dimensionality of the flow equations does not increase when one evolves additional observables.

\subsection{\label{sec:magnus2}Magnus(2) Approximation}

Having illustrated the advantages of the Magnus expansion in a simple model, we would now like to apply it in many-body calculations. Unlike in coupled cluster theory where the BCH formula for the similarity transformed Hamiltonian terminates at finite order, Eqs.~\ref{eq:magnusderiv} and~\ref{eq:bch} involve an infinite-order series of nested commutators that generate up to $A$-body operators. To make progress, we introduce the Magnus(2) truncation in which all commutators (as well as $\Omega(s),\eta(s)$ and $H(s)$) are truncated to the NO2B level. Even with this approximation, the expressions for $d\Omega/ds$ and $H(s)$ involve an infinite number of terms. However, for both Eqs.~\ref{eq:magnusderiv} and~\ref{eq:bch} at the NO2B level, we empirically find that the magnitude of terms decreases monotonically in $k$ for all systems studied thus far. Therefore, we numerically truncate Eqs.~\ref{eq:magnusderiv} at the $k^{\rm th}$ term if 
\begin{equation}
\label{eq:magnusconv}
\biggl|\frac{B_k\|ad_{\Omega}^k(\eta)\|}{k!\|\Omega\|}\biggr|< \epsilon_{\rm deriv}\,.
\end{equation}

For the truncation of \eqref{eq:bch}, we could use a similar criteria as for the derivative expression. However, since we are interested in the ground-state energy, we use a simpler condition where the series is truncated when the zero-body piece of the $k^{\rm th}$ term falls below some threshold, 
\begin{equation}
\label{eq:bchconv}
\biggl|\frac{\{ad^k_{\Omega}(H)\}_{0b}}{k!}\biggr| < \epsilon_{\rm BCH}\,.
\end{equation}
In the calculations presented below, we will find that the final results are insensitive to large variations in $\epsilon_{\rm deriv}$ and $\epsilon_{\rm BCH}$, which we take as an {\it a posteriori} justification for our truncations.

\

\section{\label{sec:implementation}Hamiltonians and Implementation}
Before presenting the results of IM-SRG(2) and Magnus(2) calculations of the homogeneous electron gas (HEG) and $^{16}$O, we review some details of our implementations for both systems. For the homogeneous electron gas, we perform our calculations for the closed-shell configuration of $N=14$ electrons in a cubic box with periodic boundary conditions. Note that if one is interested in extrapolating to the thermodynamic limit, calculations should be done for a larger closed-shell configurations of $N=38,54,66,\ldots$ electrons, with finite-size corrections for the kinetic and potential energy taken into account. Here we neglect these corrections since our primary purpose is to demonstrate the effectiveness of the Magnus expansion, and the quasi-exact Full Configuration Interaction Quantum Monte Carlo (FCIQMC) results we compare against also neglect these corrections~\cite{booth2012}. The relevant single particle orbitals are plane waves with quantized momenta
\begin{equation}
\psi_{\bold{k}\sigma}(\bold{r}) = \frac{1}{\sqrt{L^3}}e^{i\bold{k}\cdot \bold{r}}\chi_{\sigma}\,,
\end{equation}
where $L^3$ is the box volume, $\chi_{\sigma}$ is a spin eigenfunction, and $\bold{k} = \frac{2\pi}{L}(n_x,n_y,n_z)$ where $n_x,n_y,$ and $n_z$ are integers. We follow common practice and use the Wigner-Seitz radius to characterize the density of the HEG,
\begin{equation}
r_s = \frac{r_0}{a_0}\,,
\end{equation}
where $a_0$ is the Bohr radius and $r_0$ is defined in terms of the inverse density as
\begin{equation}
\frac{4}{3}\pi r_0^3 = \frac{L^3}{N}\,. 
\end{equation}
We use a basis set truncation which keeps $M$ single particle states with energy less than some cutoff $E_c$, although other choices are possible~\cite{Hagen:2013yba}. 

In the plane wave basis, the kinetic energy matrix elements are diagonal
\begin{equation}
T_{i,j} = \frac{1}{2}\bold{k_i}^2\delta_{ij}\,,
\end{equation}
and the Coulomb matrix elements are given by
\begin{equation}
\label{eq:coulme}
V_{ijkl} =\frac{1}{L^3}\frac{1}{q^2}\delta_{\sigma_i,\sigma_k}\delta_{\sigma_j,\sigma_l}\delta_{\bold{q},\bold{k}_i-\bold{k}_k}\delta_{\bold{q},\bold{k}_l-\bold{k}_j}.
\end{equation}
Note that the $\bold{q}=0$ term is omitted due to its cancellation against the inert, uniform positively charged background that is needed to make the system charge neutral~\cite{Fetter:1971aa}. Since we are interested primarily in the correlation energy, we have omitted the Madelung term in all of our calculations.

For the calculations of $^{16}$O, our starting point is the intrinsic nuclear $A$-body Hamiltonian
\begin{equation}
H = \biggl(1-\frac{1}{A}\biggr)T +T^{(2)} +V^{(2)}\,,
\end{equation}
where $T^{(2)}$ is the two-body part of the intrinsic kinetic energy, and we restrict our attention to two-nucleon interactions only. Results are presented for input NN interactions derived from the N$^3$LO (500 MeV) potential of Entem and Machleidt~\cite{Entem:2003ft} at several different free-space SRG resolution scales, $\lambda=2.0,2.7$, and $3.0$ fm$^{-1}$.

For both systems, the Magnus(2) and IM-SRG(2) calculations start by normal ordering the Hamiltonian with respect to the HF ground state. In the case of the HEG, translational invariance implies the HF orbitals are plane waves. Therefore, the HF reference state is just a Slater determinant comprised of the lowest energy doubly occupied plane wave states for $N=14$ electrons. For $^{16}$O, we must self-consistently solve the Hartree-Fock equations by expanding the unknown HF orbitals in a harmonic oscillator basis truncated to oscillator states obeying $2n+l \leq e_{\rm max}$, where $e_{\rm max}$ is sufficiently large so that the results are approximately independent of the $\hbar\omega$ value of the underlying oscillator basis. For the NN interactions used in the present calculations, a cutoff of $e_{\rm max}=8$ is sufficiently large for most purposes. Once a converged HF ground-state is obtained, the
Hamiltonian is normal-ordered w.r.t. to this solution, and the resulting in-medium zero-, one-, and two-body
operators serve as the initial values for the Magnus(2) and IM-SRG(2) flow equations. These are subsequently
integrated until sufficient decoupling is achieved, as determined by the size of the second-order many-body perturbation theory MBPT(2) contribution of the flowing Hamiltonian $H(s)$ to the ground state energy. We use a threshold of $10^{-6}$ Hartree (MeV) for the HEG ($^{16}$O) calculations, respectively, which corresponds to relative changes in the
flowing ground-state energy of $10^{-7}$ or less for both systems. 
\begin{figure*}[t]
 \centering
 \includegraphics[width=0.49\textwidth]
 {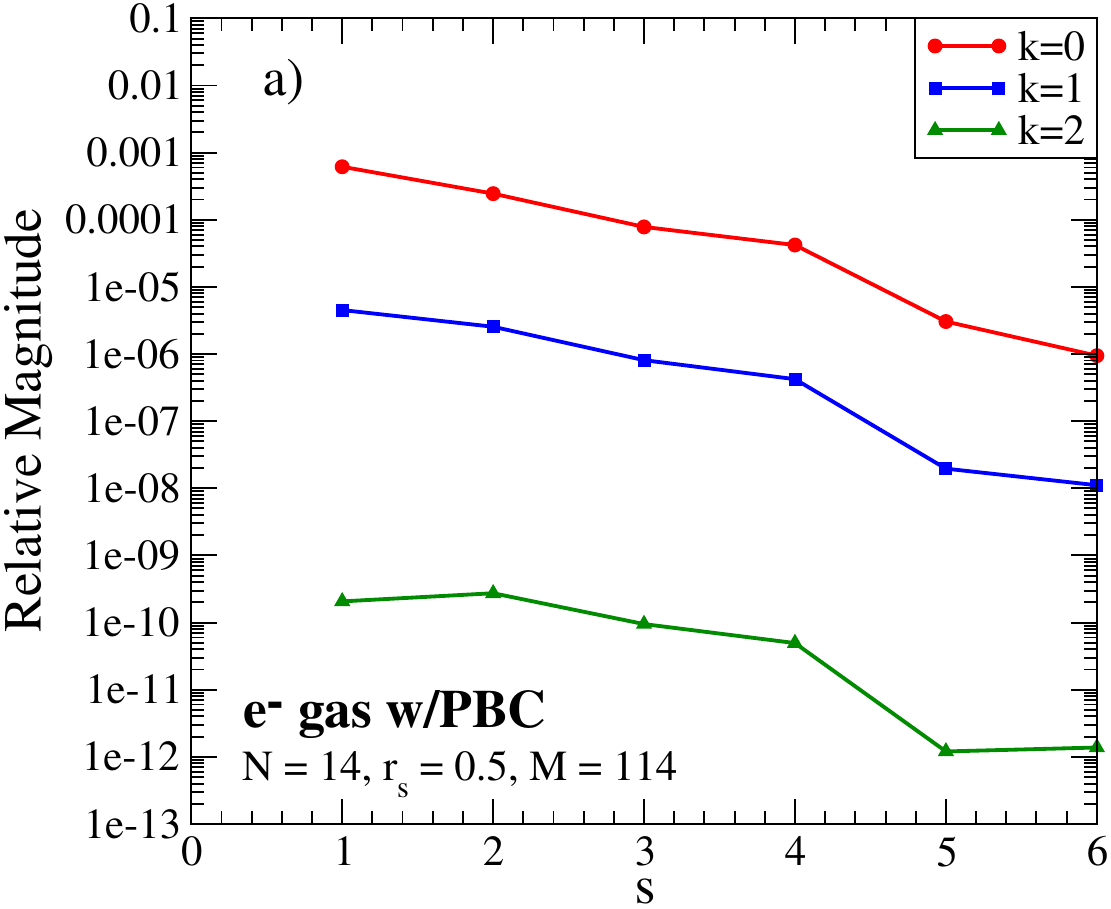}~~~%
 \includegraphics[width=0.49\textwidth]%
 {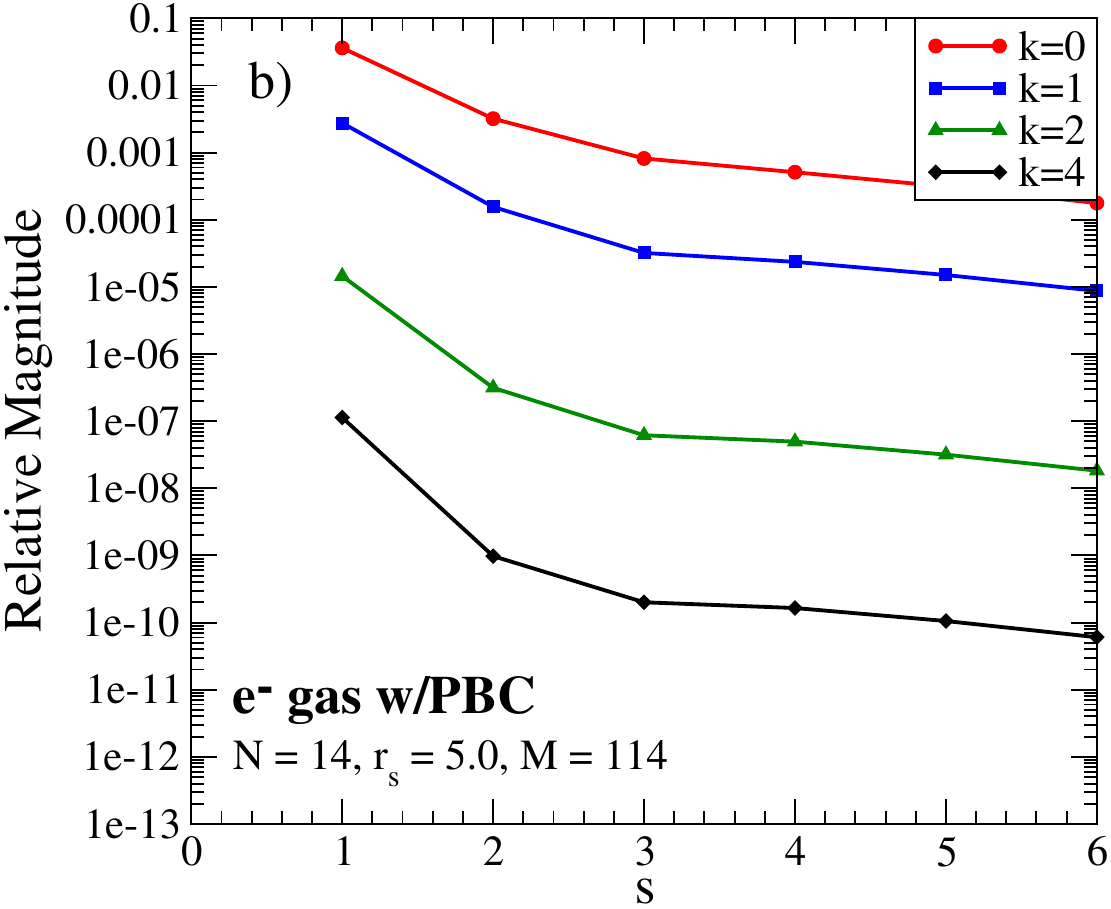}
 \includegraphics[width=0.49\textwidth]
 {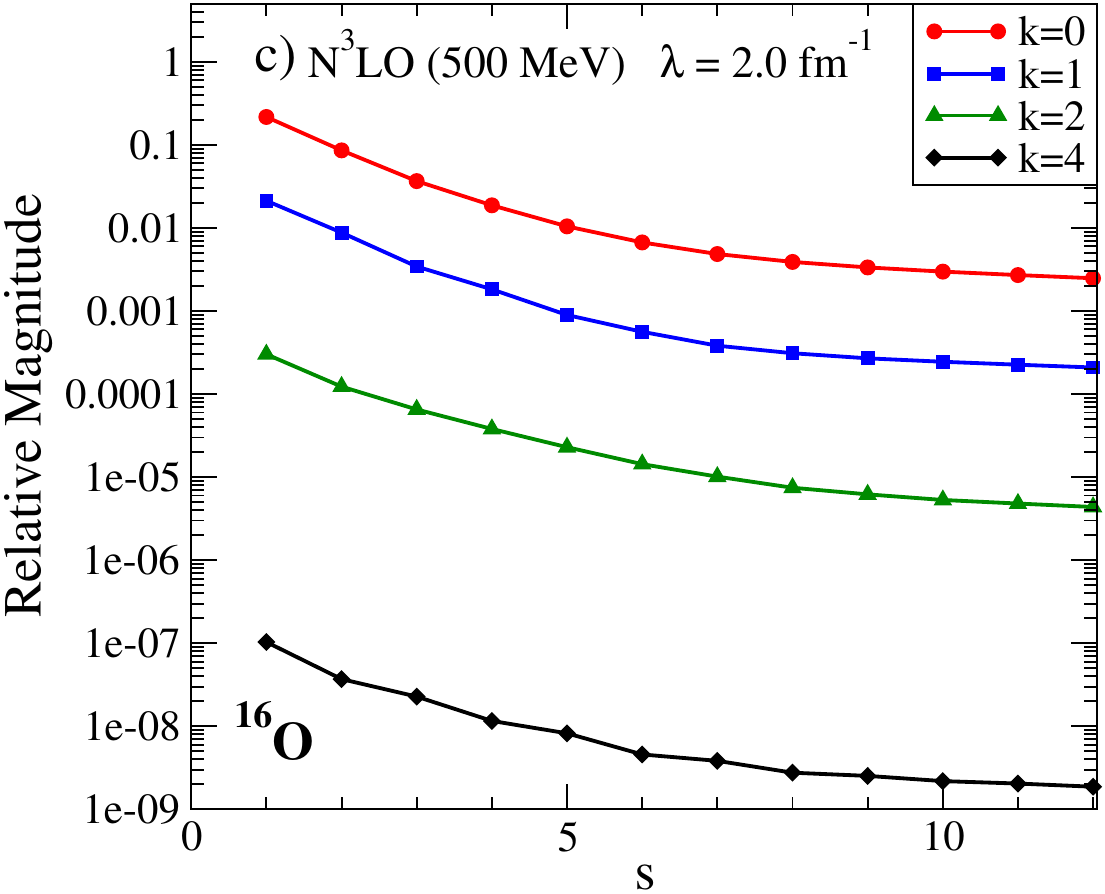}~~~%
 \includegraphics[width=0.49\textwidth]%
 {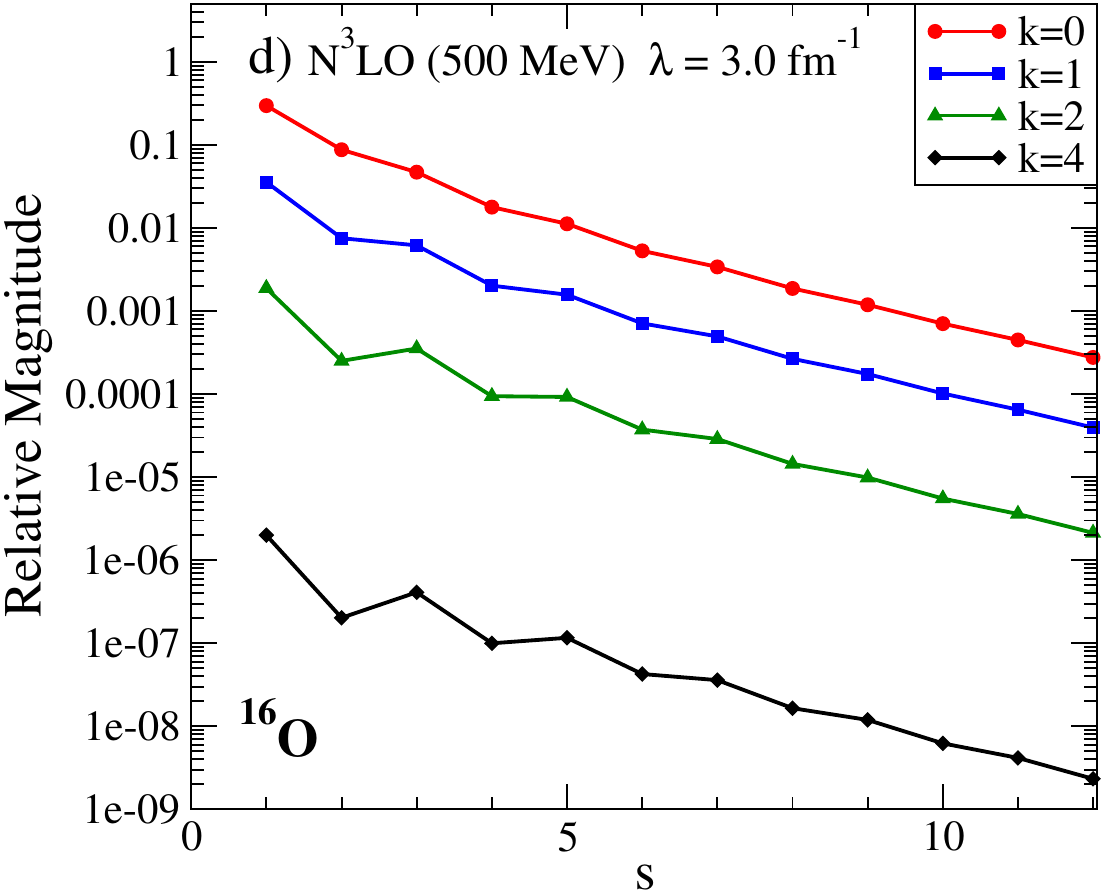}
 \caption{(Color online) Relative importance of the $k^{\rm th}$ term in the Magnus derivative as defined by the lefthand side of  Eq.~\ref{eq:magnusconv} evaluated in the NO2B approximation. The top row is for the homogeneous electron gas at Wigner-Seitz radii of a) $r_s=0.5$ and b) $r_s=5.0$. The bottom row is for $^{16}$O, starting from the chiral NN potential of Entem and Machleidt~\cite{Entem:2003ft}, softened by a free-space SRG evolution to (c) $\lambda = 2.0$ fm$^{-1}$ and (d) $\lambda = 3.0$ fm$^{-1}$. The electron gas calculations were done for $N=14$ electrons in a periodic box with $M=114$ single particle orbitals. The $^{16}$O calculations were done in an $e_{max}=8$ model space, with $\hbar\omega = 24$ MeV for the underlying harmonic oscillator basis.}
\label{fig:magnusimportance}
\end{figure*}

\section{\label{sec:results}Results}

We begin by examining the numerical evidence for truncating Eqs.~\ref{eq:magnusderiv} and~\ref{eq:bch} by hand. In Figure~\ref{fig:magnusimportance}, we plot the lefthand side of Eq.~\ref{eq:magnusconv} for the HEG (top row) and $^{16}$O (bottom row) as a function of the flow parameter. To assess the role of correlations, the HEG calculations were performed at two different Wigner-Seitz radii, $r_s=0.5$ and $r_s=5.0$, and the $^{16}$O calculations were done using NN interactions at two different resolution scales, $\lambda = 2.0$ fm$^{-1}$ and $\lambda=3.0$ fm$^{-1}$. For the HEG, the $r_s=0.5$ contributions are completely negligible  by the $k=2$ term, which is not surprising since the kinetic energy dominates in this weakly correlated high-density regime~\cite{Fetter:1971aa}. Even for the $r_s = 5.0$ case, where correlations and non-perturbative effects start to become sizable, one finds that the successive terms in Eq.~\ref{eq:magnusderiv} decrease monotonically, though the individual terms are substantially larger than for the $r_s=0.5$ case.  Analogous results are found for $^{16}$O; the individual terms are larger for the harder $\lambda=3.0$ fm$^{-1}$ interaction since the system is more strongly correlated, but they systematically decrease with increasing order $k$. 

\begin{figure*}[t!]
 \centering
 \includegraphics[width=0.49\textwidth]
 {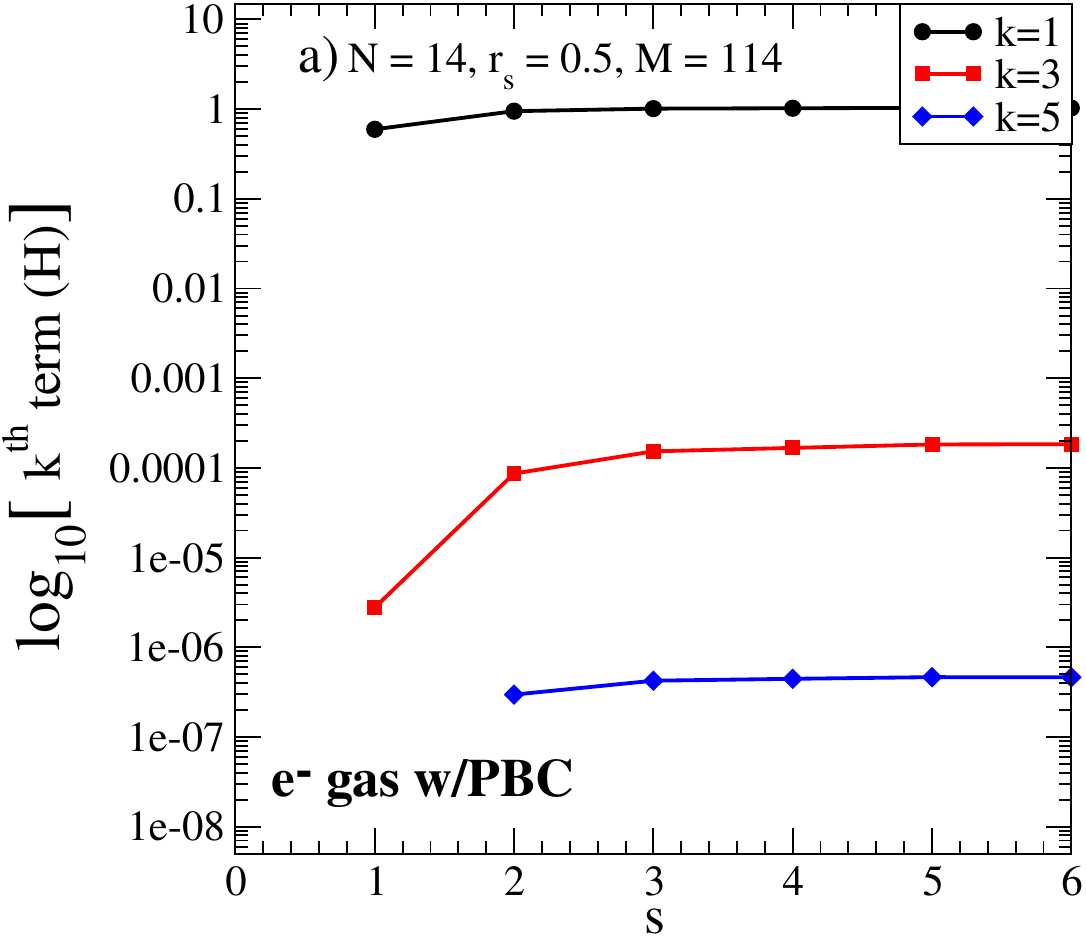}~~~%
 \includegraphics[width=0.49\textwidth]%
 {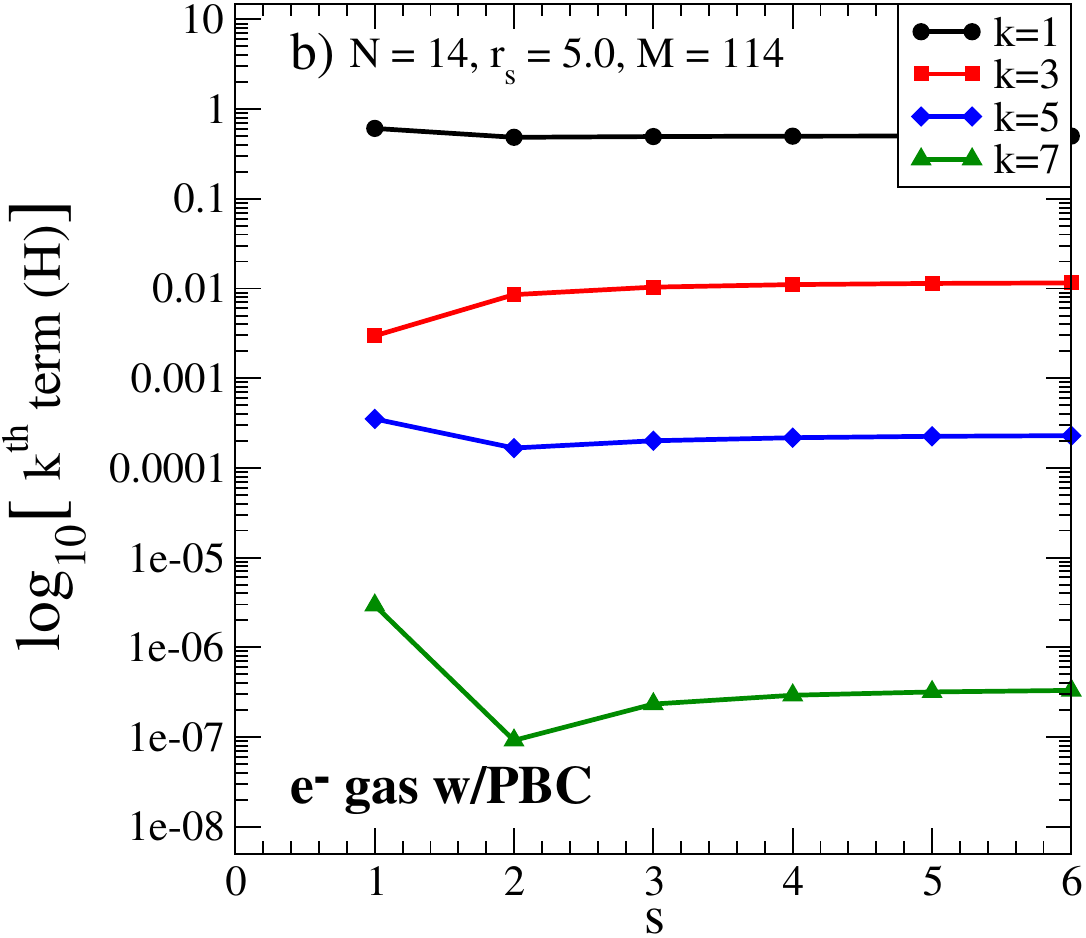}
 \includegraphics[width=0.49\textwidth]
 {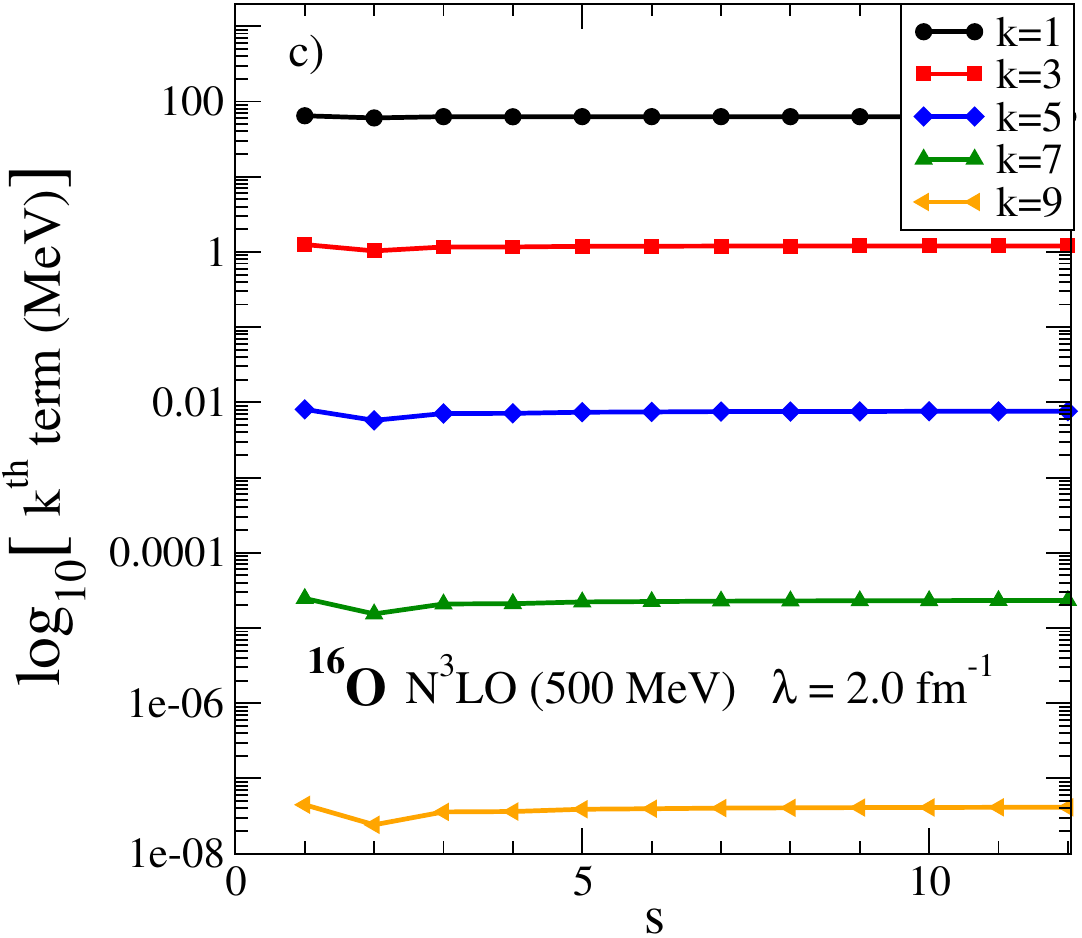}~~~%
 \includegraphics[width=0.49\textwidth]%
 {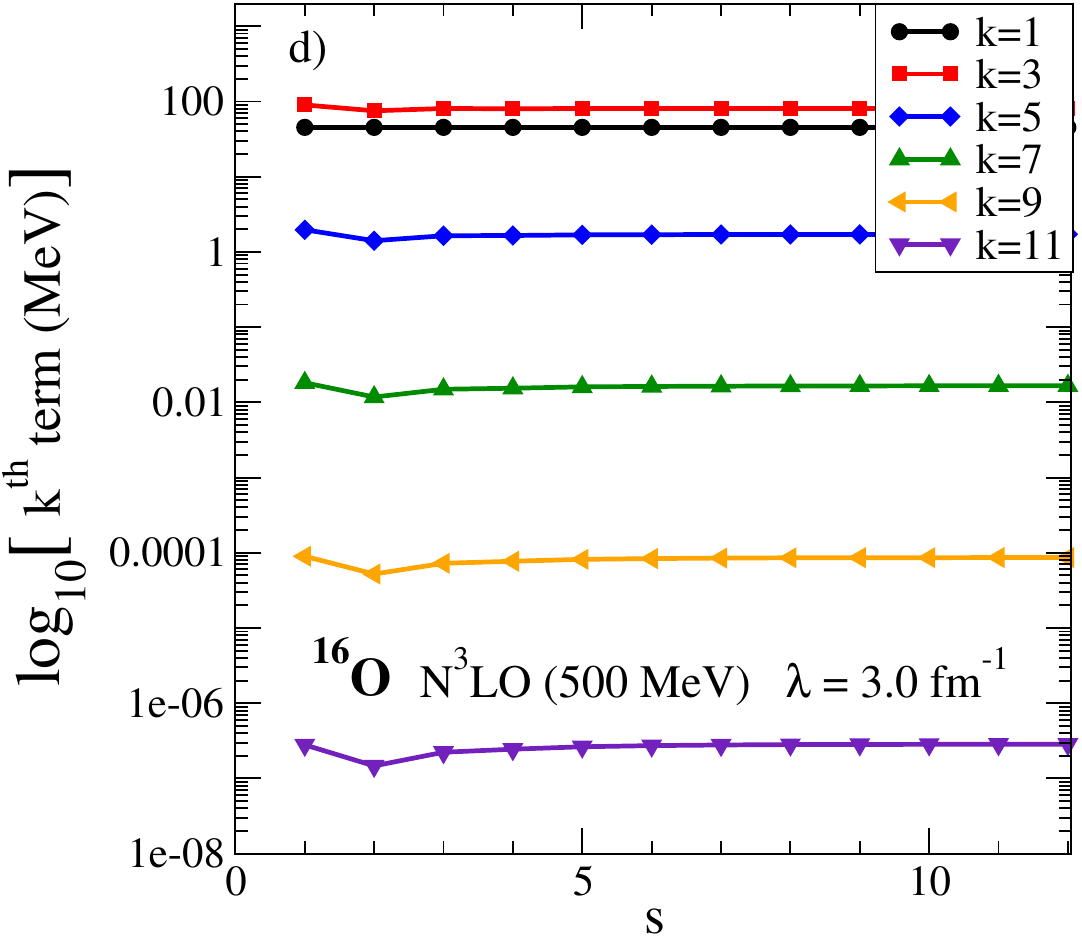}
 \caption{(Color online)  Magnitude of the 0-body contributions of the $k^{\rm th}$ term in Eq.~\ref{eq:bch} evaluated in the NO2B approximation. The top row is for the electron gas at Wigner-Seitz radii of (a) $r_s = 0.5$ and (b) $r_s=5.0$. The bottom row is for $^{16}$O, starting from the chiral NN potential of Entem and Machleidt~\cite{Entem:2003ft}, softened by a free-space SRG evolution to (c) $\lambda = 2.0$ fm$^{-1}$ and (d) $\lambda = 3.0$ fm$^{-1}$. The electron gas calculations were done for $N=14$ electrons in a periodic box with $M=114$ single particle orbitals. The $^{16}$O calculations were done in an $e_{max}=8$ model space, with $\hbar\omega = 24$ MeV for the underlying harmonic oscillator basis.}
\label{fig:bchimportance}
\end{figure*}

Figure~\ref{fig:bchimportance} tells a similar story for the BCH formula, where the lefthand side of Eq.~\ref{eq:bchconv} is plotted as a function of the flow parameter. In all cases, we see the importance of successive terms decreases monotonically.  Reassuringly, we find that the final results in our calculations are essentially independent of the convergence criteria provided $\epsilon_{{\rm deriv}} \lesssim 10^{-4}$ and $\epsilon_{{\rm BCH}} \lesssim 10^{-4}$, where the latter is in units of Hartree (MeV) for the HEG ($^{16}$O) calculations, respectively. For instance, raising both convergence criteria from $10^{-8}$ to $10^{-4}$ changes the ground state energy at the 1 eV ($10^{-7}$ Hartree) level in the $^{16}$O (HEG) calculations, respectively.

\begin{figure}[t]
  \includegraphics[width=.98\columnwidth]{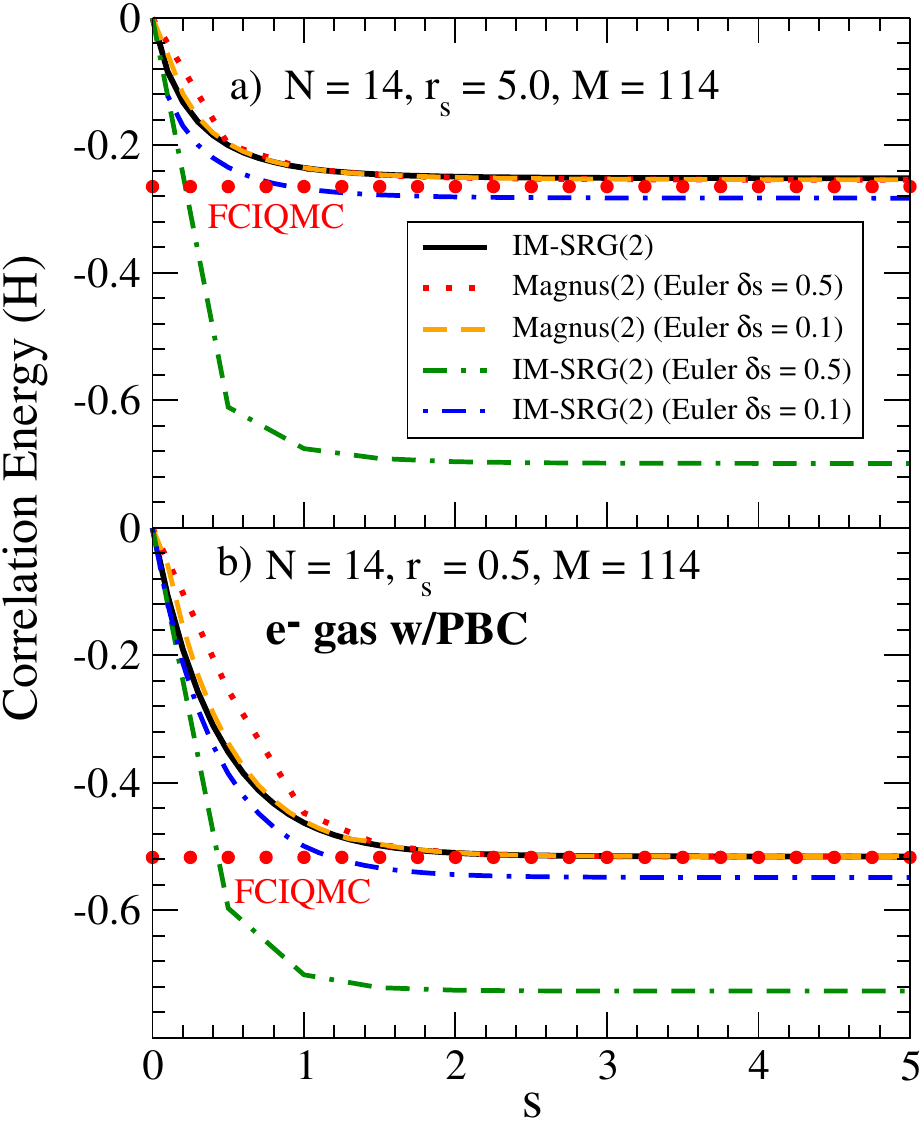}
  \caption{\label{fig:timestep_HEG}
   (Color online) Flowing IM-SRG(2) and Magnus(2) HEG correlation energy, $E_0(s)-E_{\rm HF}$, for Wigner-Seitz radii of a) $r_s=5.0$ and b) $r_s=0.5$. The solid black line corresponds to IM-SRG(2) results using an adaptive solver based on the Adams-Bashforth method, while the other lines correspond to Magnus(2) and IM-SRG(2) results using different Euler step sizes. The red circles denote the quasi-exact FCIQMC results of Ref.~\cite{booth2012}.}
\end{figure}

\begin{figure}[t]
  \includegraphics[width=.98\columnwidth]{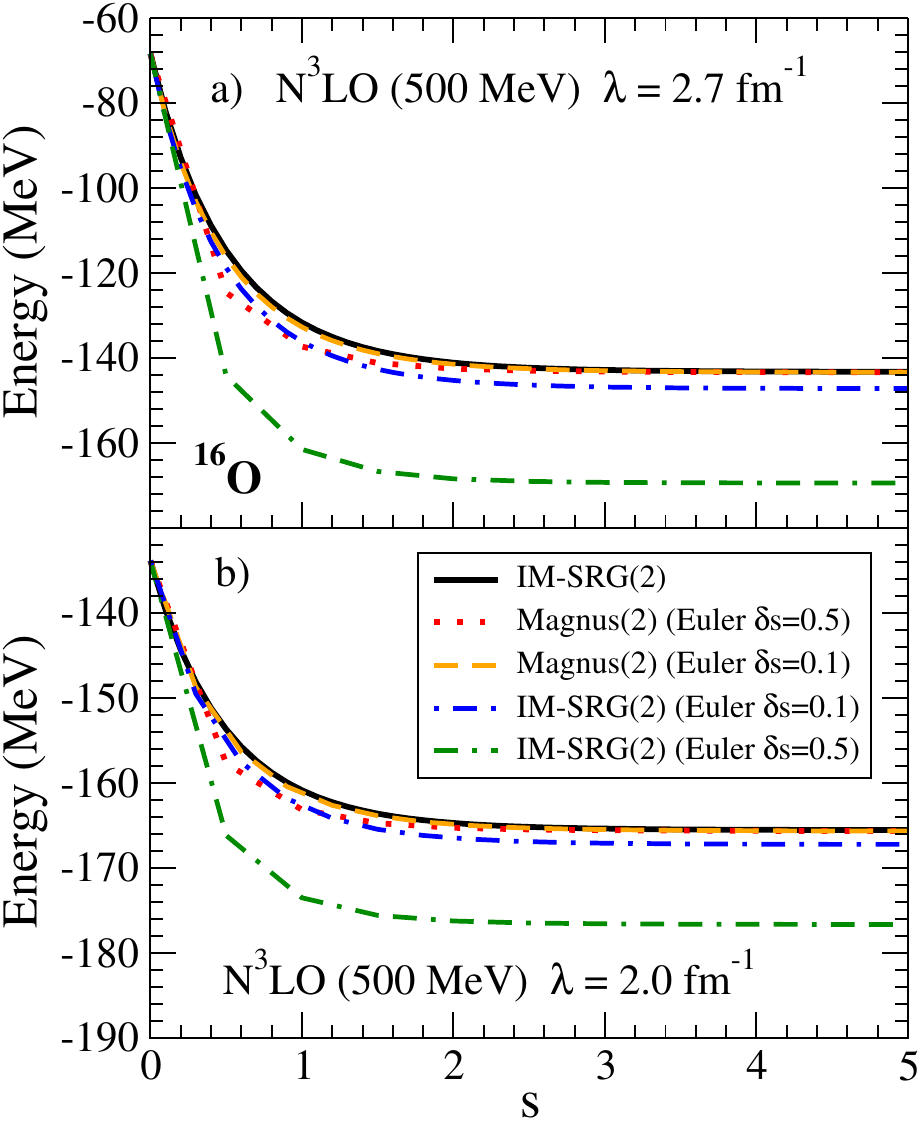}
  \caption{\label{fig:timestep_O16}
   (Color online) Flowing IM-SRG(2) and Magnus(2) ground state energy, $E_0(s)$, for $^{16}$O starting from the N$^3$LO NN interaction of Entem and Machleidt~\cite{Entem:2003ft} evolved by the free-space SRG to a) $\lambda=2.7$ fm$^{-1}$ and $\lambda = 2.0$ fm$^{-1}$. The solid black line corresponds to IM-SRG(2) results using an adaptive solver based on the Adams-Bashforth method, while the other lines correspond to Magnus(2) and IM-SRG(2) results using different Euler step sizes. The calculations were done in an $e_{max}=8$ model space, with $\hbar\omega = 24$ MeV for the underlying harmonic oscillator basis.}
\end{figure}

 As was illustrated for the toy model in Section~\ref{sec:magnusform}, the key advantage of the Magnus expansion is that one can use a first-order Euler method to accurately solve the flow equations. We now demonstrate that the same conclusion holds for realistic IM-SRG calculations. Referring to Figs.~\ref{fig:timestep_HEG} and~\ref{fig:timestep_O16}, we show the 0-body part of the flowing Hamiltonian $H(s)$ versus the flow parameter for the electron gas\footnote{For the HEG, we plot $E_0(s)-E_{HF}$, which approaches the correlation energy at large $s$. } and $^{16}$O. The black solid lines denote the results of a standard IM-SRG(2) calculation using the adaptive ODE solver of Shampine and Gordon, while the other curves denote IM-SRG(2) and Magnus(2) calculations using a first-order Euler method with different step sizes $\delta s$. For the electron gas, the exact FCIQMC results~\cite{booth2012} are shown for reference. Unsurprisingly, the IM-SRG(2) Euler calculations are very poor, with the various step sizes converging to different large-$s$ limits. The Magnus(2) calculations, on the other hand, converge to the same large-$s$ limit in excellent agreement with the standard IM-SRG(2) results. The insensitivity to step size is due to the fact that the time step errors accumulate in $\Omega(s)$ as opposed to $H(s)$. At the end of the flow, $\Omega(s)$ is still an anti-hermitian operator, and the transformation in Eq.~\ref{eq:bch} is unitary, up to truncation errors in the NO2B approximation.   

Given that the Magnus(2) results are independent of step size over the range considered, one might try to keep increasing the step size to reach the ground state in fewer steps. This unfortunately is not possible, as the flow tends to diverge with too large of a time step.  One of the strengths of the SRG approach is that the transformation is adapted as the Hamiltonian is transformed. With too large of a time step, we rob the method of this flexibility and run the risk of applying a ``large rotation'' of the Hamiltonian that induces large three- and higher-body components.  This would not be a problem if we evaluated the BCH and Magnus derivative expressions without approximation; the method would find its way back since the large rotation is still unitary if no truncations are made.  However, in the Magnus(2) approximation we make, the neglect of the induced three- and higher-body terms can lead to a lack of convergence.  Empirically, we find that this difficulty is avoided by enforcing that at each time step the ``off-diagonal'' matrix norm $\|H^{\rm od}\|$ is decreasing.  This can be implemented by using a simple mid-point integrator algorithm and decreasing the time step if $\|H^{\rm od}\|$ has increased between the first and second half of a step.  

As a final demonstration of the utility of the Magnus expansion, we turn to the evolution of operators other than the Hamiltonian. In the conventional approach based on the direct integration of Eq.~\ref{eq:srg1}, the dimensionality of the flow equations increases with each additional operator to be evolved. In contrast, in the Magnus expansion the dimensionality of the flow equations does not change; the additional computational expense shows up only in the evaluation of the BCH formula for the transformed operator, Eq.~\ref{eq:bchop}. For a given operator $O$, we have
\begin{equation}
\langle \Psi_0|O|\Psi_0\rangle = \lim_{s\rightarrow\infty}\langle\Phi|e^{\Omega(s)}Oe^{-\Omega(s)}|\Phi\rangle\,,
\end{equation}
where $|\Phi\rangle$ is the reference state. Therefore, the 0-body piece of the transformed operator approaches the interacting ground state expectation value in the large-$s$ limit. 

\begin{figure}[t]
 \includegraphics[width=.98\columnwidth]{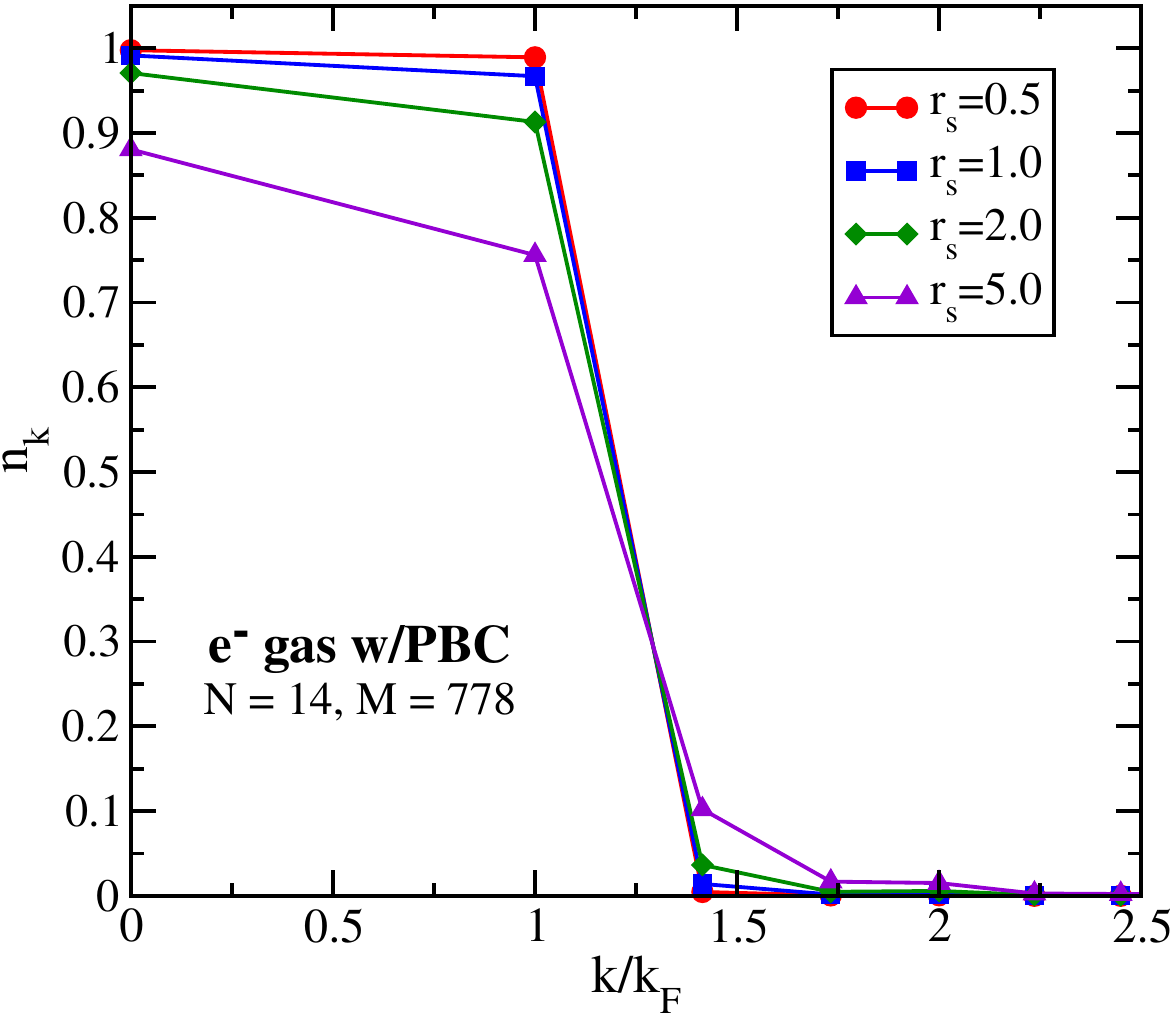}  
  \caption{\label{fig:momdist}(Color online)    (Color online) Electron gas momentum distributions calculated in the Magnus(2) approximation. The calculations were done for $N=14$ electrons in a periodic box with $M=778$ single particle orbitals. }
\end{figure}

\begin{figure}[t]
\includegraphics[width=.98\columnwidth]{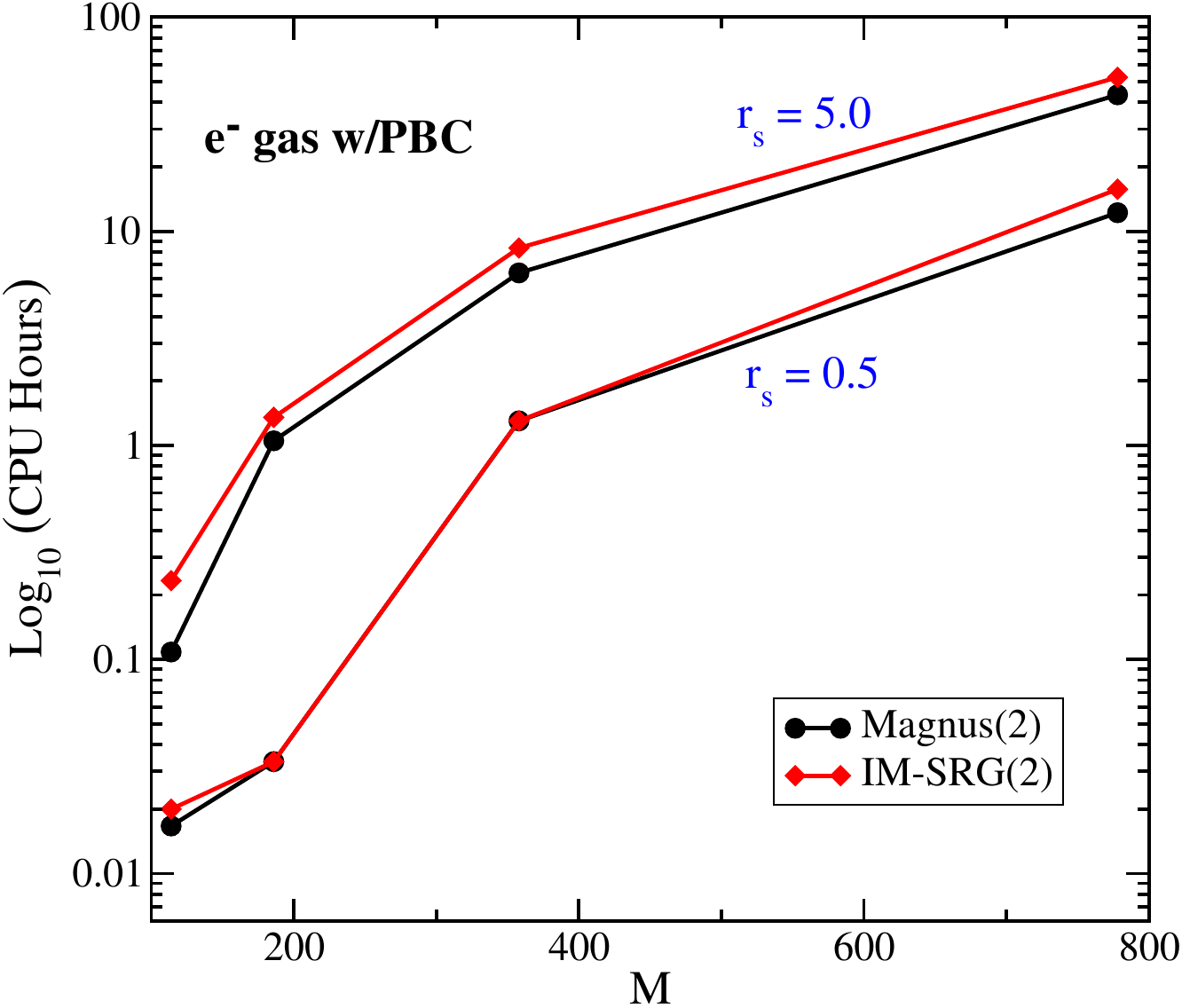}  
\caption{\label{fig:timing}(Color online)  Timing for Magnus(2) and IM-SRG(2) HEG calculations as the single particle basis is increased.  The two bottom curves are for $r_s=.5$ and the top for $r_s=5$. For the Magnus(2) timing, this includes the calculations of the momentum distributions.
}
\end{figure}

As a proof-of-principle, we perform a Magnus(2) evolution to evaluate the ground state expectation value of the momentum distribution operator $\hat{n}_{\bf k}\equiv a^{\dagger}_{\bf k}a_{\bf k}$ for the HEG, and the generalized center of mass (COM) Hamiltonian for the $^{16}$O nucleus,
\begin{equation}
\label{eq:comham}
H_{\rm cm}(\tilde{\omega}) = T_{\rm cm} + \frac{1}{2}mA\tilde{\omega}^2R^2_{\rm cm} - \frac{3}{2}\hbar\tilde{\omega}\,.
\end{equation}
Figure~\ref{fig:momdist} shows the Magnus(2) ground state momentum distribution for a system of $N=14$ electrons in a periodic box for several different Wigner-Seitz radii. Even with the neglect of finite size corrections and the extremely coarse momentum grid due to the small box sizes considered, the qualitative behavior agrees with expectations for the electron gas; correlations become more important at larger $r_s$, leading to a stronger depletion of modes with $k<k_F$ and smaller discontinuity at the Fermi surface. We note that the Magnus(2) results are in good agreement with the IM-SRG(2) calculations based on Eq.~\ref{eq:srg1} as well as results generated by the Feynman-Hellman theorem, but at a fraction of the cost. In addition to providing a memory-efficient means for evolving operators beyond the Hamiltonian, Fig.~\ref{fig:timing} shows that the Magnus(2) approximation gives a small but robust computational speedup for a range of basis sets, even including the additional effort of generating the momentum distributions, which were not computed in the IM-SRG(2) timings. 


For our second illustration of operator evolution, we consider the generalized COM Hamiltonian, Eq.~\ref{eq:comham}. In calculations of nuclei, the ground state expectation value of this quantity is useful to diagnose whether approximate solutions of the Schr\"odinger equation are contaminated by spurious COM motion. Since nuclei are self-bound objects governed by a translationally-invariant Hamiltonian, an exact solution of the Schr\"odinger equation must factorize into the product of a wave function for the physically relevant intrinsic motion times a wave function for the COM coordinate,
\begin{equation}
\label{eq:comfac}
|\Psi\rangle = |\psi\rangle_{\rm in}\otimes|\psi\rangle_{\rm cm}\,.
\end{equation}
\begin{figure}[t]

  \includegraphics[width=.98\columnwidth]{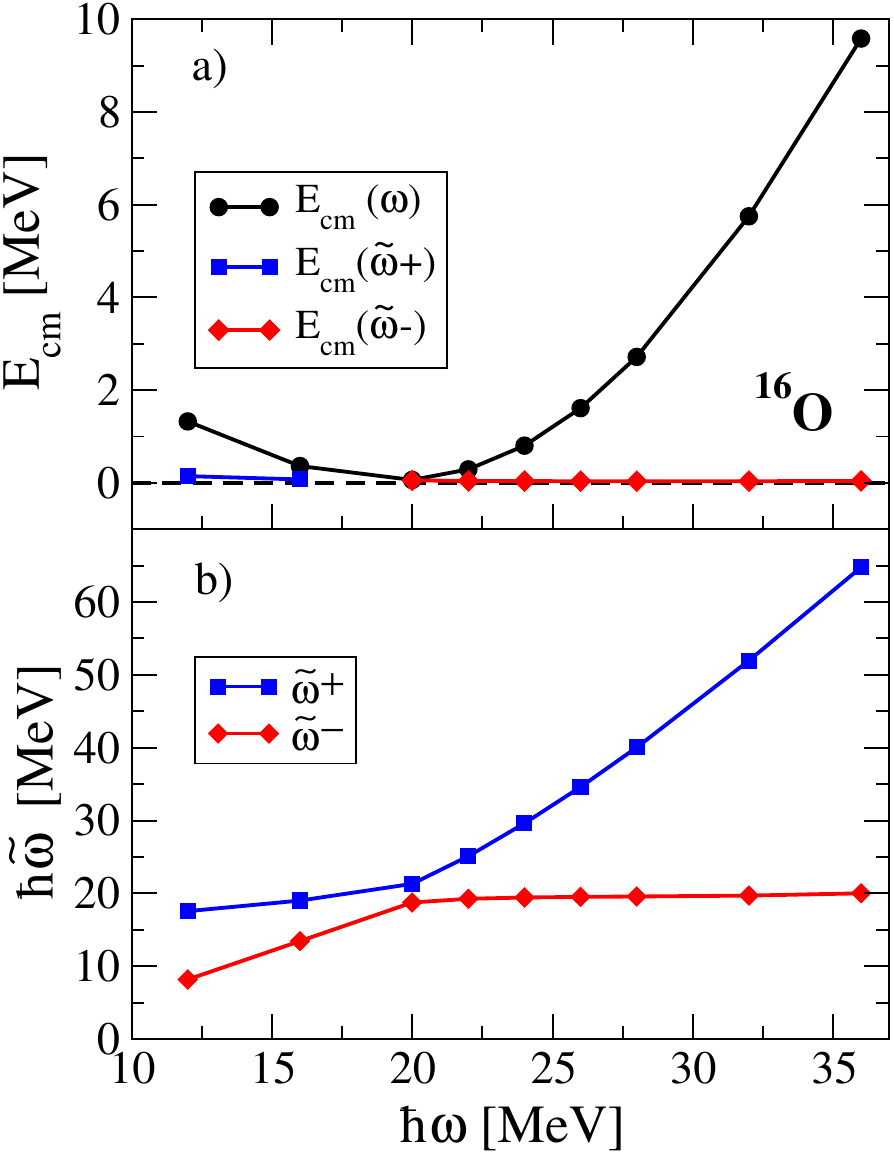}  
  \caption{\label{fig:comfac}(Color online)    (Color online) Center of mass diagnostics for Magnus(2) calculations of $^{16}$O starting from the N$^{3}$LO NN interaction of Entem and Machleidt~\cite{Entem:2003ft} evolved by the free-space SRG to $\lambda=2.0$ fm$^{-1}$. See the text for details. The calculations were done in an $e_{\rm max}=9$ model space.}
\end{figure}

As is well known, there are two strategies to rigorously guarantee this factorization; one can work in a translationally-invariant basis from the outset, or one can work in a so-called full $N\hbar\omega$ model space comprised of all $A$-particle harmonic oscillator Slater determinants with excitations up to and including $N\hbar\omega$. Neither choice is optimal since the former is limited to light nuclei due to the factorial scaling of the required antisymmetrization, while the latter limits the choice of the single particle orbitals to the harmonic oscillator basis and doesn't carry over to methods that are capable of reaching heavier nuclei, such as coupled cluster theory and the IM-SRG where it is more natural to define the model space via an energy cutoff (e.g., $2n+1 \leq e_{\rm max}$) on the single particle states. In the case of calculations with an $e_{\rm max}$ cutoff, there is no analytical guarantee that the COM and intrinsic wave functions factorize.

In Ref.~\cite{Hagen:2009pq}, Hagen and collaborators gave an ingenious prescription to diagnose whether or not Eq.~\ref{eq:comfac} is satisfied in such calculations. The basic idea is to assume that the factorized COM wave function is a Gaussian, and is therefore the ground state of $H_{\rm cm}(\tilde{\omega})$ with eigenvalue zero. Note that $\tilde{\omega}\neq \omega$ in general, where $\omega$ is the frequency of the underlying oscillator basis. The prescription to find $\tilde{\omega}$ involves solving a quadratic equation

\begin{equation}
\label{eq:omegatilde}
\hbar\tilde{\omega} = \hbar\omega +\frac{2}{3}E_{\rm cm}(\omega) \pm \sqrt{\frac{4}{9}(E_{\rm cm}(\omega))^2 +\frac{4}{3}\hbar\omega E_{\rm cm}(\omega)},
\end{equation}
where
\begin{align}
E_{\rm cm}(\omega)&\equiv \langle\Psi|H_{\rm cm}(\omega)|\Psi\rangle \\
&=\lim_{s\rightarrow\infty}\langle\Phi|e^{\Omega(s)}H_{\rm cm}(\omega)e^{-\Omega(s)}|\Phi\rangle\\
 &= \lim_{s\rightarrow\infty}\bigl\{e^{\Omega(s)}H_{\rm cm}(\omega)e^{-\Omega(s)}\bigr\}_{\rm 0b}\,.
\end{align}

Since there are two roots of Eq.~\ref{eq:omegatilde}, we choose the one that gives a smaller value for $E_{\rm cm}(\tilde{\omega})$. Applying this prescription to our calculations of $^{16}$O, we obtain the results shown in Fig.~\ref{fig:comfac}. In the top panel, we see that the expectation value of the COM Hamiltonian $H_{\rm cm}(\omega)$ is approximately zero for $\omega\approx 20$ MeV, but varies parabolically and becomes rather large away from this point. This suggests that if Eq.~\ref{eq:comfac} is satisfied, the frequency of the factorized COM Gaussian should have $\tilde{\omega}\approx 20$ MeV. This is born out in the bottom panel, where the two roots of Eq.~\ref{eq:omegatilde} are plotted as a function of  $\hbar\omega$. Picking the root that minimizes $E_{\rm cm}(\tilde{\omega})$, we find that indeed $\tilde{\omega}\approx 20$ MeV over a wide range of $\omega$, and that $E_{\rm cm}(\tilde{\omega})\approx 0$.  Since the excitation energy for the first spurious COM mode is $\hbar\tilde{\omega}\approx 20$ MeV, while $E_{\rm cm}(\tilde{\omega})$ ranges between $0.03$-$0.14$ MeV over the entire range of $\hbar\omega$, we conclude that the factorization of COM/intrinsic motion is satisfied to an excellent approximation.

\section{Summary and Outlook \label{sec:conclusions}}
In this paper, we have shown how the Magnus expansion can be used to bypass computational limitations arising from the large memory demands of high-order ODE solvers typically used in IM-SRG calculations. The success of the Magnus expansion derives from the fact that by reformulating Eq.~\ref{eq:srg1} as a flow equation for the operator $\Omega(s)$, where $U(s) = e^{\Omega(s)}$, we are able to use a simple first-order Euler ODE solver without any loss of accuracy, resulting in substantial memory savings and a modest reduction in CPU time. In conventional formulations of the SRG, time step errors accumulate directly in the evolved $H(s)$, necessitating the use of a high-order solver to preserve an acceptable level of accuracy.  In the Magnus expansion, even though sizable time step errors accumulate in $\Omega(s)$ with a first-order Euler method, upon exponentiation the transformation is still unitary, and the transformed $H(s)=U^{\dagger}(s)HU(s)$ is unitarily equivalent to the initial Hamiltonian modulo any truncations made (e.g., the NO2B approximation or numerical truncation of the infinite series) in evaluating the BCH formula. 

After introducing the basic formalism of the Magnus expansion and illustrating its numerical virtues for a schematic matrix model, we turned to realistic many-body calculations of the homogeneous electron gas (HEG) and $^{16}$O.  To make progress, we introduced the Magnus(2) approximation in which all operators ($H(s), \Omega(s), \eta(s)$) and commutators are truncated at the NO2B approximation, and the non-terminating Magnus derivative, Eq.~\ref{eq:magnusderiv}, and BCH formula, Eq.~\ref{eq:bch}, are truncated numerically. In all cases studied, our calculations converge to a final answer that is independent of step size and agrees well with IM-SRG(2) calculations using a high-order solver. Moreover, our final results are independent of the precise convergence criteria for truncating  Eqs.~\ref{eq:magnusderiv} and~\ref{eq:bch} provided that $\epsilon_{\rm deriv}\lesssim 10^{-4}$ and $\epsilon_{\rm BCH}\lesssim10^{-4}$ MeV (Hartree) for the $^{16}$O (HEG) calculations.

The evaluation of observables besides the Hamiltonian poses considerable challenges in the IM-SRG since the evolution of each additional observable roughly doubles the dimensionality of the flow equations. In the Magnus expansion formulation, the evolution of additional operators is trivial since one solves flow equations for the unitary transformation and then constructs the evolved observables by application of the BCH formula. The dimensionality of the flow equations is fixed, regardless of how many additional operators are being evolved. Proof-of-principle operator evolutions were carried out using the Magnus expansion for the momentum distribution in the HEG, and the generalized COM Hamiltonian in $^{16}$O. This opens the door for the ab-initio calculation of a variety of properties (radii, transition matrix elements, response functions, etc.) in addition to energies for medium-mass nuclei.

\section*{Acknowledgements}
We thank Morten Hjorth-Jensen and Heiko Hergert for useful discussions. This work was supported in part by the National Science Foundation under Grant No. PHY-1404159, and by the NUCLEI SciDac Collaboration under DOE Grant No. DE-SC000851.
\newpage
\hyphenation{Post-Script Sprin-ger}

\end{document}